\documentclass[journal]{IEEEtran}
\IEEEoverridecommandlockouts
\usepackage{cite}
\usepackage{amsmath,amssymb,amsfonts}
\usepackage{cleveref}
\usepackage{algorithm}
\usepackage{algpseudocode}
\algrenewcommand\algorithmicrequire{\textbf{Initialize:}}
\usepackage{graphicx}
\usepackage{subcaption}
\usepackage{textcomp}
\usepackage{xcolor}
\usepackage{array}
\definecolor{EmphOrange}{RGB}{220,120,0}
\newcolumntype{C}[1]{>{\centering\arraybackslash}p{#1}}
\usepackage[subrefformat=parens]{subcaption}
\def\BibTeX{{\rm B\kern-.05em{\sc i\kern-.025em b}\kern-.08em
    T\kern-.1667em\lower.7ex\hbox{E}\kern-.125emX}}
\begin{document}

\title{Design and Implementation of Schwarz Information Criterion-Aided Intelligent Decentralized Resource Allocation in
Dynamic LoRa Networks
}

\author{
\IEEEauthorblockN{Aohan Li, ~\IEEEmembership{Senior Member,~IEEE}, Ryota Ariyoshi, Mikio Hasegawa,~\IEEEmembership{ Member,~IEEE},\\ 
Miao Pan,~\IEEEmembership{Senior Member,~IEEE}, 
Tomoaki Ohtsuki,~\IEEEmembership{Senior Member,~IEEE}, and Zhu Han,~\IEEEmembership{Fellow,~IEEE}}
%\thanks{Manuscript received XX XX, 2026; revised XX XX, 2026. (Corresponding Author: Aohan Li)}
\thanks{
Aohan Li and Ryota Ariyoshi are with the Department of Computer and Network Engineering, The University of Electro-Communications, Tokyo, Japan %and the Department of Electrical Engineering, Tokyo University of Science, Tokyo, Japan. 
(Email: 
aohanli@ieee.org, a2431010@gl.cc.uec.ac.jp)}
\thanks{Mikio Hasegawa is with the Department of Electrical Engineering, Tokyo University of Science, Tokyo, Japan.(Email:hasegawa@ee.kagu.tus.ac.jp)}
\thanks{Miao Pan and Zhu Han are with the Department of Electrical and Computer Engineering, University of Houston, Houston, TX, 77204, USA. (Email: mpan2@uh.edu, hanzhu22@gmail.com)}
\thanks{Tomoaki Ohtsuki is with the Department of Information and Computer Science, Keio University, Yokohama, Japan. (Email: ohtsuki@keio.jp)}
}

\maketitle

\begin{abstract}
This paper proposes a lightweight distributed learning method for selecting transmission parameters in Long-Range (LoRa) networks that adapts to dynamically changing communication environments.  
In the proposed method, the Thompson Sampling (TS) is adopted for transmission parameter selection, whereas the Schwarz Information Criterion (SIC) is employed for environmental change detection.
TS is a reinforcement learning approach that effectively balances exploration and exploitation by updating parameters based on probability distributions. Additionally, it demonstrates stable performance even with a small number of trials, thereby making it well-suited for LoRa end devices (EDs) with limited memory capacity and computational resources.
Furthermore, to address the issue that TS-based methods strongly depend on past learning histories and therefore adapt slowly to abrupt changes in communication environments, a statistical change detection mechanism based on the SIC is integrated into our proposed method. 
SIC is adopted because it can detect environmental changes with low computational cost and is suitable for implementation on resource-constrained LoRa EDs.
When a change in the communication environment is detected by SIC, the learning history of TS is reset, thereby enabling rapid re-learning under new environmental conditions.
Moreover, to achieve fully distributed communication parameter selection while enhancing transmission reliability and energy efficiency, the proposed method relies solely on Acknowledgment (ACK) feedback and the selected transmission parameters.
Experimental results demonstrate that the proposed method improves the transmission success rate from 64.0\% to 71.1\% and increases energy efficiency from 293.9 bit/J to 328.3 bit/J compared with the conventional Upper Confidence Bound (UCB)1-tuned scheme under high-density dynamic LoRa networks.
\end{abstract}

\begin{IEEEkeywords}
LoRa, Decentralized resource allocation, Energy efficiency, Reinforcement learning, Schwarz information criterion, Multi-arm bandit, Thompson sampling theorem
\end{IEEEkeywords}

\section{Introduction}
In recent years, with the development of the Internet of Things (IoT) technologies, a large number of IoT devices have been densely deployed in various applications such as smart cities, agricultural monitoring, environmental sensing, and infrastructure management \cite{b1}. Since many IoT devices operate with limited battery capacity, maintaining reliable communication under strict energy constraints is essential. 
Long-Range (LoRa) is one of the representative Low-Power Wide-Area Network (LPWAN) technologies and is widely used as a low-power long-range wireless communication system based on Chirp Spread Spectrum (CSS) modulation \cite{b2, b5}. It enables long-distance communication over several kilometers and supports multi-year battery operation, making it suitable for large-scale IoT deployments \cite{b6}. LoRa operates in unlicensed bands and adopts an Aloha-based MAC protocol; therefore, in dense networks, packet collisions and interference become severe, significantly degrading both transmission success rate and energy efficiency \cite{b7,b8,b9}. In addition, LoRa communication performance strongly depends on transmission parameter configurations such as channel (CH), transmission power (TP), and bandwidth (BW), and inappropriate parameter settings can drastically reduce communication reliability and energy efficiency \cite{b3,b4,b14}.

Therefore, an efficient transmission parameter configuration mechanism is essential to maintain high communication reliability and energy efficiency in high-density LoRa networks. 
To address this challenge, various transmission parameter optimization methods for LoRa networks have been proposed and can generally be classified into centralized \cite{d1,d2,d3,d5,d6,b11} and distributed approaches \cite{d7,d8,d9,d10,b13,b15,b25}.
In centralized approaches, a network server centrally manages the communication conditions of all devices and assigns appropriate transmission parameters to each device to improve overall network performance. However, such approaches suffer from scalability limitations due to increased server-side processing loads. 
In addition, in LoRa networks, transmission parameters are typically configured through downlink control messages from the gateway. However, in LoRa networks, downlink communication opportunities are limited and subject to duty-cycle constraints. As a result, waiting for control messages to configure transmission parameters introduces additional latency and energy consumption, making centralized approaches less suitable for large-scale networks \cite{b11}.
In contrast, decentralized approaches allow each device to autonomously determine its transmission parameters without receiving transmission parameter configuration messages from the gateway, thereby improving scalability and reducing both energy consumption and communication resource usage.

As representative decentralized approaches, methods based on the Multi-Armed Bandit (MAB) have been extensively studied for adaptive Resource Allocation (RA), owing to their low computational complexity and ease of implementation on IoT devices \cite{d10,b13,b15,b16,b17,b18,b25,d7,d8}. 
However, most of them did not consider energy efficiency \cite{d10,b13,b15,b16,b17,b18}.
Among the related work that considered energy efficiency \cite{b25,d7,d8}, Upper Confidence Bound (UCB)1-tuned is well known for its efficient exploration capability based on upper confidence bounds. However, because the UCB-based methods strongly depend on accumulated historical statistics, outdated observations can adversely affect decision-making in dynamic environments, resulting in delayed adaptation after environmental changes.
To improve adaptability under such dynamic environments, a recent study has incorporated statistical change detection mechanisms based on the Schwarz Information Criterion (SIC) into MAB frameworks \cite{b19}.
However, as this method depends on spectrum sensing over all channels to detect changes in the channel environment, its application in practical real-time IoT systems remains challenging. To overcome this challenge, our proposed method performs change detection solely using  Acknowledgment (ACK)  histories already available during normal communication, eliminating the need for additional sensing operations.

On the other hand, optimization methods based on Thompson Sampling (TS) have also been presented for solving sequential decision-making problems under uncertainty 
\cite{b23,b24}. TS performs probabilistic sampling from posterior distributions, which makes arms with higher expected rewards more likely to be selected even in the early learning stage, achieving a favorable balance between exploration and exploitation. Compared with the UCB-based methods, TS does not require exhaustive initial exploration over all parameter combinations, which can not only improve early-stage transmission success rate and energy efficiency, but also accelerate convergence. These properties are particularly effective in scenarios with a large number of parameter combinations or where rapid re-convergence is required after environmental changes.

Motivated by the above description, this paper proposes a distributed RA method that integrates TS with SIC-based change detection to enable rapid adaptation to dynamic environmental conditions in LoRa networks.
In the proposed method, each LoRa End Device (ED) learns transmission parameters using ACK feedback and transmission energy consumption.
Furthermore, environmental changes are detected using SIC based on ACK histories of the selected transmission channels, and the learning history is reset when a change is detected.
The main contributions of this paper are summarized as follows.
\begin{itemize}
    \item
    We propose a lightweight decentralized RA method based on TS. The proposed method requires maintaining only a small number of statistical variables\textcolor{EmphOrange}{,} and does not involve complex matrix operations or centralized control. As a result, it incurs low computational and memory overhead, thereby making it well suited for implementation on memory-constrained IoT devices.
     \item
    The proposed method incorporates energy consumption-related parameters and successfully transmitted payload into the reward function, thereby achieving simultaneous improvements in successful transmission rate and energy efficiency.
    \item 
    By incorporating an SIC-based change detection principle, the proposed method ensures adaptability to dynamically varying wireless environments. Furthermore, reinforcement learning based on TS enables efficient learning without requiring exhaustive exploration of all parameter combinations. This enables rapid adaptation while mitigating performance degradation in the early learning stage as well as after environmental changes.
    \item Experimental evaluations using real LoRa devices confirm that the proposed method significantly outperforms conventional UCB1-tuned and SIC-enhanced UCB1-tuned in terms of both transmission success rate and energy efficiency under dynamically changing environments.
\end{itemize}

This paper is structured as follows. Section II reviews related work. Section III describes the system model and problem formulation. Section IV presents the proposed method. Section V provides the implementation and performance evaluation. Finally, Section VI concludes the paper.

\section{Related Work}
In this section, existing studies related to RA for LoRa networks are reviewed. First, the centralized RA methods \cite{d1, d2, d3, d5, d6, b11} are described. Next, the decentralized RA approaches \cite{d7,d8,d9,d10,b13,b15,b25} are introduced. 
A comparative summary of the related research in RA for LoRa networks is presented in Table~\ref{tab:related_work}.
\begin{table*}[t]
\centering
\caption{Comparison of RA Methods in LoRa Networks}
\label{tab:related_work}
\renewcommand{\arraystretch}{1.2}
\begin{tabular}{|C{2.2cm}|C{2.2cm}|C{2.6cm}|C{1.8cm}|C{2.0cm}|C{2.0cm}|C{2.0cm}|}
\hline
\textbf{Reference} &
\textbf{Centralized / Distributed} &
\textbf{Selected Transmission Parameters} &
\textbf{EE Consideration} &
\textbf{Computational Complexity} &
\textbf{Dynamic Environment Detection} &
\textbf{Real-Device Experiment} \\
\hline
\cite{d6} & Centralized & SF & $\checkmark$ & High & -- & -- \\
\hline
\cite{d3} & Centralized & SF, TP & $\checkmark$ & High & -- & -- \\
\hline
\cite{d1} & Centralized & CH, SF, TP & $\checkmark$ & High & -- & -- \\
\hline
\cite{b11} & Centralized & CH, SF, TP, CR & $\checkmark$ & High & -- & -- \\
\hline
\cite{d2} & Centralized & CH, SF, TP & $\checkmark$ & High & -- & $\checkmark$ \\
\hline
%\cite{d4} & Centralized & CH & -- & High & $\checkmark$ & -- \\
%\hline
\cite{d5} & Centralized & CH, TP & $\checkmark$ & High & $\checkmark$ & $\checkmark$ \\
\hline
\cite{d9} & Distributed & SF, TP & $\checkmark$ & High & -- & -- \\
\hline
\cite{d10} & Distributed & CH, SF, TP & -- & Low & -- & -- \\
\hline
\cite{b13} & Distributed & CH, SF & -- & Low & -- & $\checkmark$ \\
\hline
\cite{b15} & Distributed & CH & -- & Low &--  & $\checkmark$ \\
\hline

\cite{d7} & Distributed & SF & $\checkmark$ & Low & -- & -- \\
\hline
\cite{d8} & Distributed & SF, TP & $\checkmark$ & Low & -- & -- \\
\hline
%\cite{b14} & Distributed & CH & -- & Low & --  & $\checkmark$ \\
%\hline
%\cite{b16} & Distributed & CH & -- & Low & -- & $\checkmark$ \\
%\hline
%\cite{b17} & Distributed & CH, SF & -- & Low &  -- & $\checkmark$ \\
%\hline
%\cite{b18} & Distributed & CH & -- & Low &  -- & $\checkmark$ \\
%\hline
\cite{b25} & Distributed & CH, TP, BW &\checkmark & Low &  -- & $\checkmark$ \\
\hline
Proposed Method & Distributed & CH, TP, BW & $\checkmark$ & Low & $\checkmark$ & $\checkmark$ \\
\hline
\end{tabular}
\end{table*}

\subsection{Centralized RA}
Centralized RA approaches have been widely studied for achieving highly optimal RA by leveraging centrally collected global network state information \cite{d1, d2, d3, d5, d6, b11}.

\cite{d6} proposed Artificial Intelligence–Empowered RA (AI-ERA), a centralized Deep Neural Network (DNN)-based SF assignment scheme that significantly improves packet success rates compared with Adaptive Data Rate (ADR) and Blind ADR (BADR).  
The capability of AI-ERA to support both static and mobile nodes demonstrates its high applicability to practical deployment scenarios.
\cite{d3} proposed a topology-aware Graph Neural Network (GNN) learning framework that models multihop LoRa networks as graph structures and optimizes SF and TP using centralized graph neural networks.  
By integrating topological information with analytical collision probability models, the proposed approach enables highly accurate energy efficiency optimization while considering retransmissions, significantly advancing RA research in multihop LoRa environments.
\cite{d1} proposed a Matching-based and Two-stage Attention-enhanced (M-TAG) Graph Convolutional Network (GCN) framework for transmission parameter optimization in multi-gateway LoRa networks. The framework employs a hierarchical centralized architecture that jointly optimizes CH, TP, and Spreading Factor (SF) by integrating matching theory and multi-agent reinforcement learning.
By incorporating GCNs and two-stage attention mechanisms, M-TAG precisely models inter-ED interference and achieves highly accurate system-wide energy efficiency maximization, positioning it as a representative state-of-the-art solution for dense multi-gateway environments.
\cite {b11} introduces a new algorithm using the MAB technique to configure the EDs’ transmission parameters, including CH, SF, TP, and coding rate. The performance of the proposed algorithm is evaluated through simulation results, which indicate that the proposed method outperforms other Adaptive Data Rate (ADR)-based methods.
\cite{d2} proposed a joint optimization framework for LoRa uplink systems that formulates the problem of ED resource block association, SF assignment, and TP control to minimize uplink TP.  
By employing heuristic allocation algorithms, the proposed approach obtains high-quality solutions with low computational complexity, and its effectiveness has been validated through experiments on real devices, demonstrating a significant reduction in power consumption. This highlights the practical applicability of centralized optimization schemes.
%\cite{d4} proposed a dynamic spectrum access framework that integrates the Listen-After-Collision (LAC) based Deep Q-Network with the improved Thompson Sampling Algorithm (DQN-iTSA) to enhance spectrum utilization efficiency under non-stationary wireless environments.
\cite{d5} proposed a centralized deep reinforcement learning framework that jointly optimizes CH selection, TP, and feedback policies under smart jamming conditions, and the proposed approach has been validated through real device experiments.

Despite their excellent performance in multi-dimensional parameter optimization, energy efficiency maximization, and high-accuracy inference, centralized approaches suffer from scalability limitations. As the number of devices increases, communication overhead increases significantly. Furthermore, EDs have to receive transmission parameters from gateways for each transmission, which results in additional energy consumption and latency.

\subsection{Decentralized RA}
To address the scalability limitations of centralized schemes, numerous distributed learning-based RA approaches have been proposed, where each ED autonomously learns and selects its transmission parameters. 
Distributed approaches offer superior scalability and low communication overhead since centralized control is not required \cite{d7,d8,d9,d10,b13,b15,b25}.

\cite{d9} proposed Multi-Agent learning for LoRa (MALoRa), an attention-based multi-agent deep reinforcement learning framework that enables cooperative learning while considering inter-ED interference relationships, thereby significantly improving system-wide energy efficiency.  
This distributed cooperative learning paradigm, which achieves global energy efficiency maximization, represents one of the most important advances among distributed learning-based approaches.
\cite{d10} proposed MAB-based distributed RA schemes for Unmanned Aerial Vehicle (UAV)-assisted LoRa networks, demonstrating that both packet delivery ratio improvement and power consumption reduction can be achieved solely through online learning. 
The fact that these methods do not require training datasets further highlights their high applicability to real-world deployment.
Furthermore, \cite{b13,b15} proposed a series of methods based on Tug-of-War (ToW) dynamics, demonstrating that extremely lightweight distributed learning relying solely on ACK information can maintain stable communication success rates even in dense and dynamic environments, as verified through real-device experiments.

\cite{d7} proposed Fast, mULti-armed bandit approach for optiMal SpreadIng allocatioN in lorA networks (FULMINA), a fully distributed SF allocation scheme based on UCB-type MAB algorithms that achieves fast convergence with low computational and memory requirements.  
By explicitly incorporating energy consumption into the reward design, the proposed approach achieves significant power reduction compared with Q-learning-based methods, demonstrating excellent characteristics from an energy efficiency optimization perspective.
\cite{d8} proposed LInk-Weight-EXP3 (LI-WEX), a distributed MAB-based scheme that incorporates link prior knowledge and jointly optimizes SF, TP, BW, and center frequency (CF).  
By reducing the exploration space and employing weighted reward designs, the proposed approach achieves fast and robust learning, representing a typical distributed method capable of multi-dimensional transmission parameter optimization.
\cite{b25} proposed an energy-efficient MAB-based distributed RA method for LoRa networks, enabling each device to independently select appropriate transmission parameters, including CH, TP, and BW. Experimental results using real LoRa devices show that the proposed method outperforms fixed allocation, ADR-Lite, and $\epsilon$-greedy methods in both transmission success rate and energy efficiency.

The above studies collectively constitute pioneering work that systematically demonstrates the feasibility of ultra-lightweight distributed learning in practical LoRa networks.
However, although distributed approaches exhibit excellent scalability, implementation simplicity, and low communication overhead, challenges remain in terms of explicitly optimizing energy efficiency and robustly adapting to rapidly changing wireless environments.

\section{System Model}
\begin{figure}
\centering
\includegraphics[width=8cm]{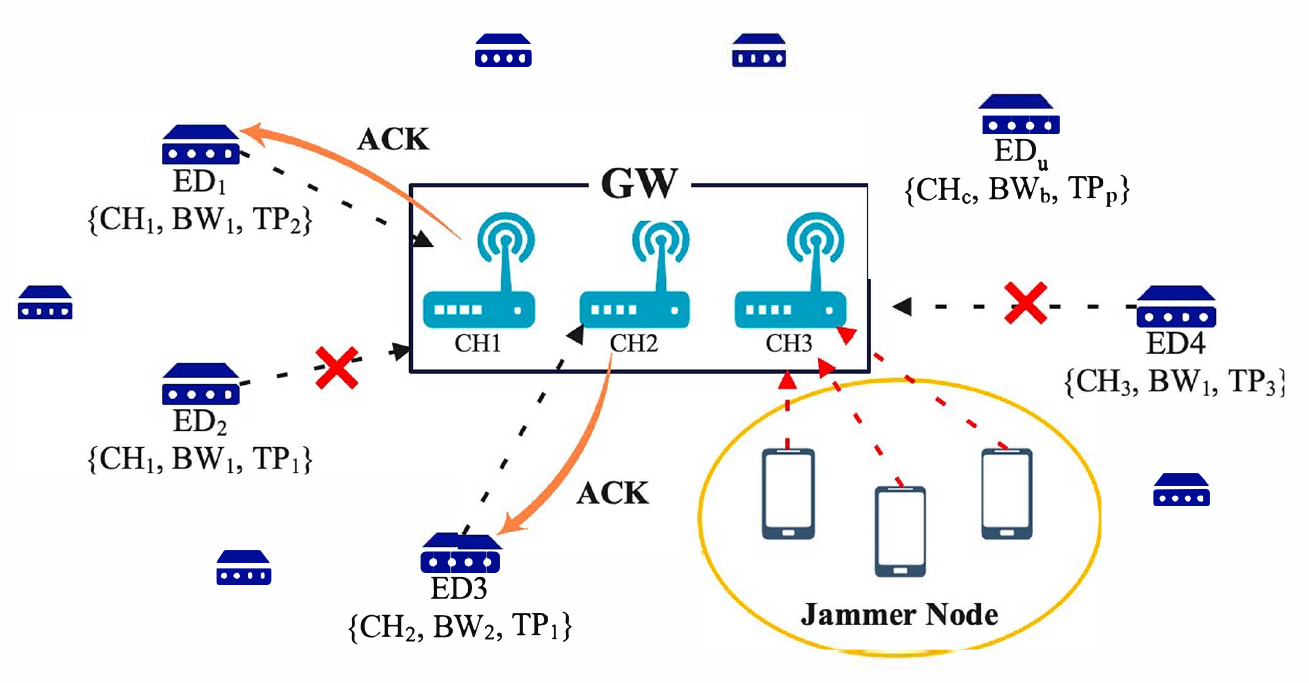}
\caption{System Model.}
\label{fig:s1}
\end{figure}

The LoRa network considered in this study is illustrated in Fig.~\ref{fig:s1}, consisting of a single gateway (GW), $N_u$ LoRa EDs, and an external interference source (jammer nodes). The set of LoRa EDs is denoted as $U = \{1, 2, \ldots, u, \ldots, N_{u}\}$. In the LoRa network, $N_c$ channels are available, which is denoted as $C = \{1, 2, \ldots, c, \ldots, N_{c}\}$. Each ED selects one channel to transmit data to the GW. However, when the jammer occupies or blocks a specific channel, transmissions over that channel are more likely to experience interference. Therefore, we assume a dynamic communication environment where the jammer intermittently affects part of the frequency spectrum, causing certain channels to transition from an available state to an unavailable state during operation. Each ED thus needs to adaptively select the appropriate transmission parameters according to the varying channel conditions.
Let $N_p$ be the number of TP levels, and $N_b$ be the number of BW options. 
The set of TP levels is denoted as $P = \{1, 2, \ldots, p, \ldots, N_{p}\}$,
and that of BW options as $B = \{1, 2, \ldots, b, \ldots, N_{b}\}$,
Let $K = \{k_{1}, k_{2}, \ldots, k_{i}, \ldots, k_{N_c \times N_p \times N_b}\}$ be the set of all possible combinations of CH, TP, and BW.
Each ED selects one parameter set from 
$K$ before transmitting data based on the proposed method.

Each ED transmits data at fixed intervals. Before transmission, it performs carrier sensing on the selected channel. If the channel is sensed as available, the ED proceeds to transmit using the selected combination of CH, TP, and BW. If the transmission is successful, the GW sends an ACK feedback to the ED, and the ED receives a positive reward. If no ACK is received, the transmission is considered failed, and the result is reflected in subsequent learning. Transmission parameter selection and learning processes are executed in a fully distributed manner on each ED. Based on ACK feedback and energy consumption information associated with the selected transmission parameters, each ED updates its learning policy to maximize both the success rate of the transmission and energy efficiency.

Furthermore, each ED records the presence or absence of ACKs for each transmission and accumulates this as a transmission history. In this paper, we introduce a mechanism that statistically detects changes in the communication environment based on this history using the SIC. When a change is detected, the past learning history is discarded, and the learning process is reinitialized to enable quick adaptation to the new environment. 
To represent the transmission outcomes quantitatively, each ED maintains $N_c$ binary observation sequences of ACK receptions corresponding to $N_c$ channels. The $c$-th binary observation sequence is expressed as follows:
\begin{equation}
    \Omega_c = \{\, s_{c,i} \mid s_{c,i} \in \{0, 1\},\, i = 1, 2, \ldots, l_{max}\,\},
\end{equation}
where $s_{c,i} = 1$ denotes a successful ACK reception in the $c$-th channel, and $s_{c,i} = 0$ denotes a failure.
$l_{\max}$ denotes the maximum length of each binary observation sequence. Historical data before $l_{\max}$ are discarded to maintain low memory overhead and real-time performance.
To analyze temporal variations in communication performance, this sequence is divided into sliding windows of length $W$, shifted by $F$ steps. 
Given the total sequence length $l$, the number of windows $D$ is expressed as:
\begin{equation}
D = \left\lfloor \frac{l + F - W}{F} \right\rfloor.
\end{equation}
Let $x_d$ denote the number of ACK successes in window $d \in \{1, 2, \ldots, D\}$,
and let $p_d$ represent the average ACK success probability within that window.
These statistics are then used to evaluate the temporal variations in transmission performance and to detect environmental changes.

The energy consumption model of a LoRa device used in this paper models the energy consumption
during data communication in active mode, which can be calculated below \cite{b6}. 

%\begin{enumerate}
    %\item \textbf{Energy Consumption in Active Mode ($E_{Active}$)}
    \begin{equation}
       E_{Active} = E_{WU} + E_{\text{proc}} + E_{\text{ToA}} + E_{\text{R}},
    \end{equation}   
    where \(E_{\text{WU}}\) represents the energy consumption during device wake-up, \(E_{\text{proc}}\) represents the energy consumption for transmission parameter selection by the microcontroller, \(E_{\text{ToA}}\) represents the energy consumption during data transmission, and \(E_{\text{R}}\) represents the energy consumption during the reception. The values of $E_{\text{WU}}$, \(E_{\text{proc}}\), and \(E_{\text{R}}\) depend on the specifications of the modules used in the device.
    \(E_{\text{ToA}}\) can be expressed as follows:
    %\item \textbf{Energy Consumption During Data Transmission (\(E_{\text{ToA}}\))}
    \begin{equation}
       E_{\text{ToA}} = (P_{\text{MCU}} + P_{\text{ToA}}) \cdot T_{\text{ToA}}
       \label{EToA},
    \end{equation}  
    where \(P_{\text{MCU}}\) is the power consumption due to the activation of the microcontroller, \(P_{\text{ToA}}\) is the power consumption during data transmission, which is determined by the selected TP.
    \(T_{\text{ToA}}\) is the total transmission duration, which can be calculated as follows:
    \begin{equation}
      T_{\text{ToA}} = T_{\text{Preamble}} + T_{\text{Payload}},
    \end{equation}
    where \(T_{\text{Preamble}}\) represents the duration required to transmit the preamble, and \(T_{\text{Payload}}\) represents the duration required to transmit the data payload. \(T_{\text{Preamble}}\) and \(T_{\text{Payload}}\) can be expressed as follows:
    \begin{equation}
      T_{\text{Preamble}} = (4.25 + N_{\text{P}}) \cdot T_{\text{Symbol}},
    \end{equation}
    \begin{equation}
      T_{\text{Payload}} = N_{\text{Payload}} \cdot T_{\text{Symbol}},
    \end{equation}
    where \(N_{\text{P}}\) is the number of preamble symbols, and \(N_{\text{Payload}}\) is the number of payload symbols.
    \(T_{\text{Symbol}}\) is the symbol duration, which can be calculated as follows:
    \begin{equation}
      T_{\text{Symbol}} = \frac{2^{SF}}{BW},
    \end{equation}
    where \(SF\) and \(BW\) are the used SF and BW when transmitting symbols.

In this study, the transmission success rate of a parameter set $k_i$ at time $t$ is defined as:
\begin{equation}
X_{k_i}(t) = \frac{R_{k_i}(t)}{N_{k_i}(t)},
\end{equation}
where $R_{k_i}(t)$ is the cumulative number of successful transmissions and $N_{k_i}(t)$ is the cumulative number of times the parameter set $k_i$ has been selected up to time $t$. This value reflects the probability of successful transmission based on the history of each parameter set.  

Based on this definition, the Energy Efficiency (EE) of a parameter set $k_i$ at time $t$ is expressed as

\begin{equation}
EE_{k_i}(t) = 
\frac{\text{Payload}_{k_i}(t) \times X_{k_i}(t)}{E_{\text{Active}}},
\label{eq:EE}
\end{equation}
where $\text{Payload}_{k_i}(t)$ denotes the payload size associated with the transmission parameter combination $k_i$ at time $t$.
In other words, $EE_{k_i}(t)$ indicates the number of successfully transmitted bits per unit of consumed energy.

The objective of this study is to maximize the cumulative energy efficiency of all EDs by optimally selecting the CH, TP, and BW under dynamic communication environments. 
The optimization problem is formulated as:

\begin{subequations}
\begin{align}
(\textbf{P1}) &\max_{k_i \in {K}} 
\sum_{u=1}^{N_u} \sum_{t=1}^{T} EE_{u,k_i}(t)\\
{\rm s.t.} \quad 
&C1: \sum_{c \in {C}} x_{u,c}(t) = 1, 
\quad \forall u,t  \\
&C2: \sum_{p \in {P}} y_{u,p}(t) = 1, 
\quad \forall u,t  \\
&C3: \sum_{b \in {B}} z_{u,b}(t) = 1, 
\quad \forall u,t  \\
&C4: x_{u,c}(t),\, y_{u,p}(t),\, z_{u,b}(t) \in \{0,1\}.
\end{align}
%\label{eq:objective_allEDs}
\end{subequations}

Here, $EE_{u,k_i}(t)$ represents the instantaneous energy efficiency of device $u$ when it selects parameter combination $k_i$ at time $t$, and $T$ denotes the total number of transmissions.
In addition, due to the practical transmission constraint of LoRa devices, each ED is required to select exactly one channel, one transmission power level, and one bandwidth option at each transmission time. 
To explicitly model this constraint, we introduce binary decision variables 
$x_{u,c}(t)$, $y_{u,p}(t)$, and $z_{u,b}(t)$, which indicate whether ED $u$ selects channel $c \in {C}$, transmission power $p \in {P}$, and bandwidth $b \in {B}$ at time $t$, respectively. This constraint is shown in 11 (e). 
Moreover, these variables should satisfy the following constraints.
That is, each ED selects exactly one channel (11b), one transmission power level (11c), and one bandwidth option (11d) at each transmission time.

\section{Proposed Method}
This paper proposes a lightweight distributed reinforcement learning method that integrates energy-efficient TS with SIC for adaptive transmission-parameter selection in dynamic LoRa networks. Each LoRa ED autonomously selects a combination of CH, TP, and BW based solely on local ACK feedback and transmission energy consumption, without relying on centralized control.
In the proposed method, the reward is defined as the achieved transmission efficiency in terms of successfully delivered payload bits per unit energy consumption (bit/J), so that both transmission reliability and energy efficiency are jointly optimized. Furthermore, an SIC-based statistical change detection mechanism is incorporated to identify significant environmental changes. When a change is detected, the learning history is reset to eliminate the adverse influence of outdated statistics and enable rapid re-learning under the new communication conditions.
By combining energy-aware TS with lightweight SIC-based change detection, the proposed method realizes fast adaptation to dynamic channel conditions while maintaining extremely low computational and memory complexity, making it suitable for implementation on resource-constrained LoRa EDs. In the sequel, we first introduce the SIC detection mechanism, followed by the decentralized RA based on TS in our proposed algorithm. Then, we present the overall algorithm. Finally, we analyze the computational complexity and memory requirements
of the proposed method.

\subsection{SIC-Based Statistical
Change Detection Mechanism}

To adapt to the non-stationarity of the LoRa communication environment, this paper introduces a statistical change detection mechanism based on SIC. Although SIC is generally employed as an information-theoretic criterion for model selection, it is applied here to quantitatively determine whether the statistical characteristics of the communication environment have changed. Since SIC evaluates the tradeoff between model fit and complexity, it can be used to compare statistical models that assume different ACK success probabilities before and after a potential change point. A significant increase in SIC indicates that a model assuming different success probabilities fits the observed ACK sequence better than a single-probability model, thereby signaling a change in the communication environment. 

In the SIC, two hypotheses are considered:  
(a) the success probability remains constant across all windows ($H_0$), and  
(b) the success probability changes at an unknown point ($H_1$).

(a) {\em Null Hypothesis (No Change) $H_0$}:  
All windows share the same ACK success probability $p$, i.e., $p_1 = p_2 = \cdots = p_D = p$.  
The SIC under $H_0$ is calculated as:
\begin{align}
\mathrm{SIC}(D) &= \log D - 2 \sum_{d=1}^{D} \log \binom{W}{x_d} \nonumber \\
&\quad - 2(Y - X)\log\left(\frac{Y - X}{Y}\right)
 - 2X\log\left(\frac{X}{Y}\right),
 \label{eq:sic_H0}
\end{align}
where $X$ is the total number of ACK successes and $Y$ is the total number of transmission attempts.

(b) {\em Alternative Hypothesis (Change Exists) $H_1$}:  
There exists a split point $j$ $(1 \le j < D)$ such that  
$p_1 = \cdots = p_j \neq p_{j+1} = \cdots = p_D$.  
The SIC under $H_1$ is expressed as:
\begin{align}
\mathrm{SIC}(j) &= 2\log D - 2 \sum_{d=1}^{D} \log \binom{W}{x_d} \nonumber \\
&\quad - 2(Y_j - X_j)\log\left(\frac{Y_j - X_j}{Y_j}\right)
 - 2X_j\log\left(\frac{X_j}{Y_j}\right) \nonumber \\
&\quad - 2(Y'_j - X'_j)\log\left(\frac{Y'_j - X'_j}{Y'_j}\right)
 - 2X'_j\log\left(\frac{X'_j}{Y'_j}\right),
 \label{eq:sic_H1}
\end{align}
where $X_j = \sum_{d=1}^{j} x_d$, $Y_j = jW$, $X'_j = X - X_j$, and $Y'_j = Y - Y_j$.  
The split point $j$ represents a potential boundary in the observation sequence at which the ACK success probability may change.  
In other words, SIC compares the statistical likelihoods before and after each candidate point $j$ to determine whether a change in the communication environment has occurred.

A change is detected when the following condition is satisfied:
\begin{equation}
\mathrm{SIC}(D) - \min_{1 \le j \le D-1} \mathrm{SIC}(j) > \eta,
\label{eq:sic_threshold}
\end{equation}
where $\eta$ is an empirically determined detection threshold. 
A smaller $\eta$ makes the detector more sensitive to environmental variations, enabling faster adaptation but potentially causing frequent resets due to false detections. In contrast, a larger $\eta$ reduces false alarms but may delay adaptation to abrupt environmental changes.
When this condition is satisfied at time $t$, the ED resets the learning statistics of all parameter combinations $k_i \in K$ used in the proposed SIC-TS-based method. Specifically, the Beta posterior parameters $\alpha_{k_i}$ and $\beta_{k_i}$ in TS, which represent the accumulated successful and failed transmission statistics, are reset to their initial values. In addition, the observation history $\Omega_c$ used for SIC computation is cleared. 

\subsection{Decentralized RA Based on TS}

In the proposed method, we employ a distributed learning scheme based on TS for transmission parameter selection in dense LoRa networks. In the proposed framework, each LoRa ED learns transmission parameters using only ACK feedback and transmission energy consumption, aiming to simultaneously maximize transmission reliability and energy efficiency.

Each transmission parameter combination $k_i \in K$, which consists of a CH, TP and BW, is regarded as an arm in a MAB problem. For each arm, a Beta posterior distribution is maintained to model the ACK success probability $\theta_{k_i}$, where $\theta_{k_i}$ is assumed to follow $\mathrm{Beta}(\alpha_{k_i},\beta_{k_i})$.
The Beta distribution is the conjugate prior of the Bernoulli distribution, where the parameters $\alpha_{k_i}$ and $\beta_{k_i}$ correspond to the accumulated numbers of successful and failed transmissions, respectively. At the initial stage, all arms are initialized with $\alpha_{k_i} = \beta_{k_i}=1$.
At each transmission time $t$, each ED independently draws a posterior sample $\tilde{\theta}_{k_i}$ from the corresponding Beta posterior distribution for all candidate parameter combinations $k_i \in K$. The sampled value $\tilde{\theta}_{k_i}$ represents a posterior sample of the success probability of $k_i$ at the current time and is used as the evaluation metric for selection. The ED then selects the transmission parameter combination $k^*$ that maximizes $\tilde{\theta}_{k_i}$.
Since this probabilistic selection is based on posterior sampling, parameter combinations with consistently high success probabilities are selected with high probability, while those with large uncertainty are also explored with non-negligible probability. As a result, the balance between exploration and exploitation is naturally maintained.
Instead of updating the learning statistics solely based on the number of successful transmissions, the proposed method introduces transmission efficiency as the learning update value. The reward at time $t$ is calculated for every transmission regardless of the ACK result and is defined as:
\begin{equation}
\Delta R = \frac{\mathrm{Payload}_{k_i}(t)}{E_{\mathrm{ToA}}(t)},
\end{equation}
where ${E_{\mathrm{ToA}}(t)}$ denotes the energy consumption during data transmission at the $t$-th decision. The reward is defined as a function of the energy consumption during data transmission for the following reason. The energy consumed during device wake-up and reception, as well as that required by the microcontroller to select transmission parameters, is determined by the specifications of the hardware modules. Because identical modules are assumed for all devices in this study, these components do not contribute to variations in energy consumption. In contrast, ${E_{\mathrm{ToA}}(t)}$ is directly related to the selected transmission parameters and thus has a direct impact on the overall energy consumption. According to the ACK result $s_t$, the parameters of the Beta posterior distribution are updated as:
\begin{equation}
\alpha_{k^*} \leftarrow \alpha_{k^*} + \Delta R \quad (s_t = 1), \qquad
\beta_{k^*} \leftarrow \beta_{k^*} + \Delta R \quad (s_t = 0).
\end{equation}
Note that, in the proposed method, the update value is weighted by the transmission efficiency $\Delta R$. Therefore, $\alpha_{k_i}$ and $\beta_{k_i}$ no longer represent strict Bernoulli counts but rather weighted pseudo-observations that reflect both transmission outcomes and energy efficiency. This allows the posterior distribution to favor parameter combinations that achieve successful transmissions with lower energy consumption.
Therefore, $\alpha_{k_i}$ is updated when an ACK success is observed, while $\beta_{k_i}$ is updated when a transmission failure is observed. Moreover, by introducing the energy-efficiency-based weight $\Delta R$ into the update, parameter combinations that achieve higher data delivery with lower energy consumption are more strongly reinforced and thus become more likely to be selected in future transmissions.

Furthermore, to take into account the transmission energy characteristics associated with BW and TP, a prior bias $w_{k_i}$ is introduced into the sampling values obtained by TS. 
The reason that $w_{k_i}$ is introduced can be summarized as follows. 
Although TS is effective for learning ACK success probabilities, it does not explicitly consider energy consumption. Thus, parameter combinations with high success rates but large energy consumption may be excessively selected. Since the essential optimization objective in LoRa networks is energy efficiency, it is necessary to explore low-energy parameter combinations from the early learning stage preferentially. For this reason, $w_{k_i}$ is introduced.
Hence, the final selection metric of the TS in this paper is defined as:
\begin{equation}
\theta_{k_i} = (1 - \gamma)\tilde{\theta}_{k_i} + \gamma w_{k_i},
\label{eq:theta}
\end{equation}
where the bias term $w_{k_i}$ is given by
\begin{equation}
w_{k_i} = \frac{E_{\max} - E_{k_i}}{E_{\max} - E_{\min}}.
\label{eq:TS_w}
\end{equation}
Here, $E_{k_i}$ denotes the transmission energy consumed when the transmission is performed using the parameter combination $k_i$. $E_{\min}$ and $E_{\max}$ represent the minimum and maximum transmission energy among all parameter combinations. The parameter $\gamma$ is a weighting factor that controls the contributions of the stochastic exploration term $\tilde{\theta}_{k_i}$ and the energy-based prior bias $w_{k_i}$; a larger $\gamma$ prioritizes parameter combinations with lower energy consumption, whereas a smaller $\gamma$ emphasizes learning based on ACK success probability. However, with the increase of the $\alpha_{k_i}$ and $\beta_{k_i}$, their ratio becomes less sensitive to changes in channel availability. Therefore, the SIC is introduced to detect variations in the communication environment.

\subsection{SIC-TS based Decentralized RA}
The overall procedure of the proposed method is summarized in Algorithm 1, which is independently executed by each ED. Each ED sequentially updates its transmission policy based on ACK feedback and transmission energy information. Through TS based on posterior distributions, parameter combinations with consistently high success probabilities are selected with high probability, while those with large uncertainty are also explored with non-negligible probability; thus, a natural balance between exploration and exploitation is maintained. Furthermore, when a change in the communication environment is detected by the SIC-based statistical change detection mechanism, the posterior distributions are reset, enabling rapid re-learning under a new environment.

\begin{algorithm}[t]
\small
\caption{Proposed Method}
\label{alg:sicts}
\begin{algorithmic}[1]
\Require
$t=0$, $\alpha_{k_i}=1$, $\beta_{k_i}=1$ for all $k_i\in K$, 
history buffers $\Omega_c=\emptyset$ for all $c\in {C}$

\While{$t<T$}
    \State Select $k^{*} \leftarrow \textbf{SIC--TS}(\{\alpha_{k_i},\beta_{k_i}\})$
    \State Transmit using $k^{*}$ and observe ACK $s_t \in \{0,1\}$
    \State Measure transmission energy consumption $E_{\mathrm{ToA}}(t)$ using (\ref{EToA})
    \State $\Delta R \leftarrow \dfrac{\mathrm{Payload}_{k^{*}}(t)}{E_{\mathrm{ToA}}(t)}$
    \If{$s_t = 1$}
        
        \State $\alpha_{k^{*}} \leftarrow \alpha_{k^{*}} + \Delta R$
    \Else
        %\State $\Delta R \leftarrow 0$
        
        \State $\beta_{k^{*}} \leftarrow \beta_{k^{*}} + \Delta R$
    \EndIf

    %\If{$s_t = 1$}
        
    %\Else
        
    %\EndIf

    \State Append $s_t$ to history $\Omega_{c(k^*)}$
    \State Compute SIC$(D)$ and SIC$(j)$ using~(\ref{eq:sic_H0})--(\ref{eq:sic_H1}) based on $\Omega_{c(k^*)}$
    \If{SIC$(D) - \min_j \mathrm{SIC}(j) > \eta$}
        \State Reset  $\alpha_{k_i} \leftarrow 1$, $\beta_{k_i} \leftarrow 1$ for all $k_i\in K$, 
        \State Reset all histories in $\Omega_c \leftarrow \emptyset$ for all $c\in {C}$
    \EndIf
    \State $t \leftarrow t + 1$
\EndWhile

\Function{SIC--TS}{$\{\alpha_{k_i},\beta_{k_i}\}$}
    \ForAll{$k_i \in K$}
        \State Sample $\tilde{\theta}_{k_i} \sim \mathrm{Beta}(\alpha_{k_i},\beta_{k_i})$
        \State Compute $w_{k_i}$ using~(\ref{eq:TS_w})
        \State  Compute $\theta_{k_i}$ using~(\ref{eq:theta})
    \EndFor
    \State \Return $k^{*} \leftarrow \arg\max_{k_i \in K} \theta_{k_i}$
\EndFunction
\end{algorithmic}
\end{algorithm}

In the proposed method, each LoRa ED autonomously and distributively learns and selects a transmission parameter set $k^* \in K$, consisting of a CH, TP, and BW, using only ACK feedback and the energy consumption of each transmission. At the beginning of the algorithm, the Beta posterior parameters of all parameter combinations $k_i \in K$ are initialized as $\alpha_{k_i}=1$ and $\beta_{k_i}=1$. In addition, for SIC-based change detection, the ACK history buffer $\Omega_c$ is initialized as an empty set for all channels $c \in {C}$.

At each transmission time $t$, the ED selects a transmission parameter set $k^*$ using the SIC--TS function (line~2). Within the SIC--TS function, for each candidate $k_i$, an evaluation value is computed based on TS using the Beta posterior distribution (lines~21--23), and the parameter set
\begin{equation}
k^* = \arg\max_{k_i \in K} \theta_{k_i}
\end{equation}
that maximizes the evaluation value is selected (line~25). The ED then transmits using the selected $k^*$ and observes the ACK result $s_t \in \{0,1\}$ (line~3). Simultaneously, the transmission energy consumption $E_{\mathrm{ToA}}(t)$ is measured using Eq.~(\ref{EToA}) (line~4).
The reward is defined as the transmission efficiency, and the Beta posterior distribution corresponding to the selected parameter set is updated according to the transmission outcome (lines~5--10). This weighted update proportional to the transmission efficiency promotes parameter sets with higher energy efficiency to be selected more frequently in future transmissions.

\begin{figure*}
    \centering
    \begin{subfigure}{0.8\linewidth}
        \centering
        \includegraphics[width=\textwidth]{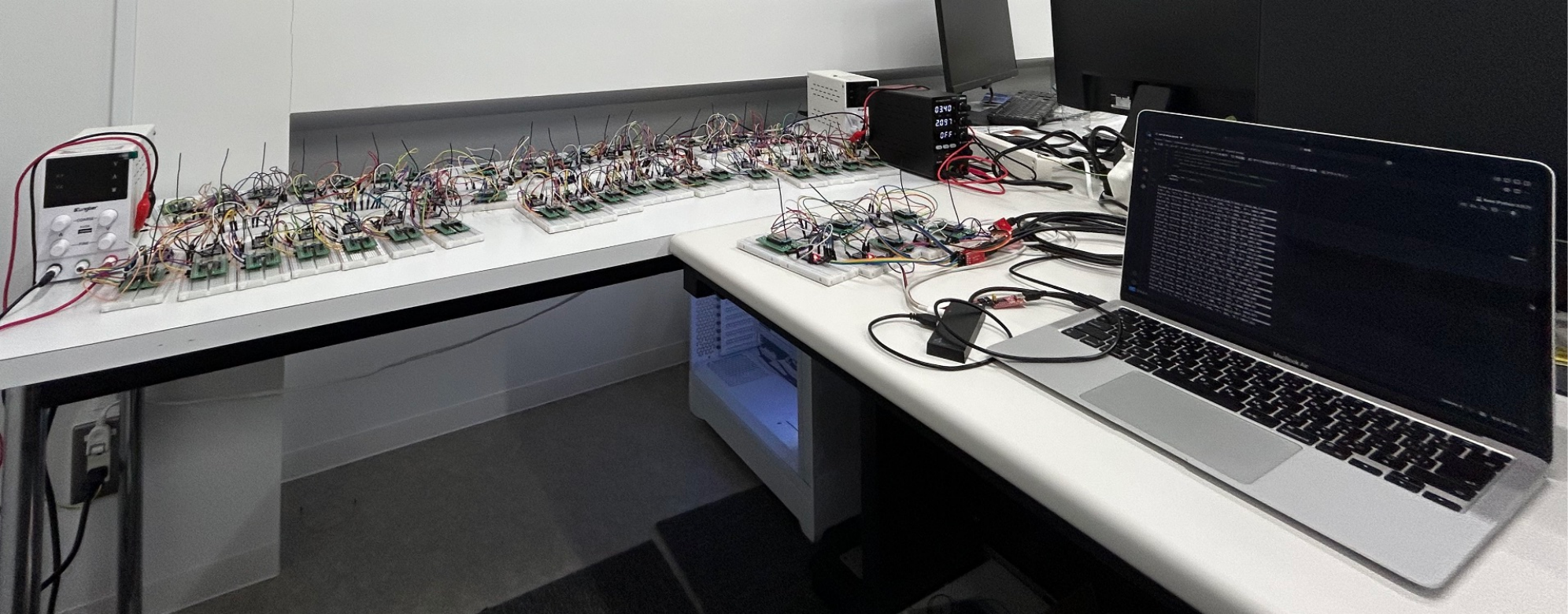}
        \caption{Overall View.}
        \label{fig:overallview}
    \end{subfigure}

    \vspace{0.5em}
    \begin{subfigure}{0.45\textwidth}
        \centering
        \includegraphics[width=\textwidth,height=3.8cm]{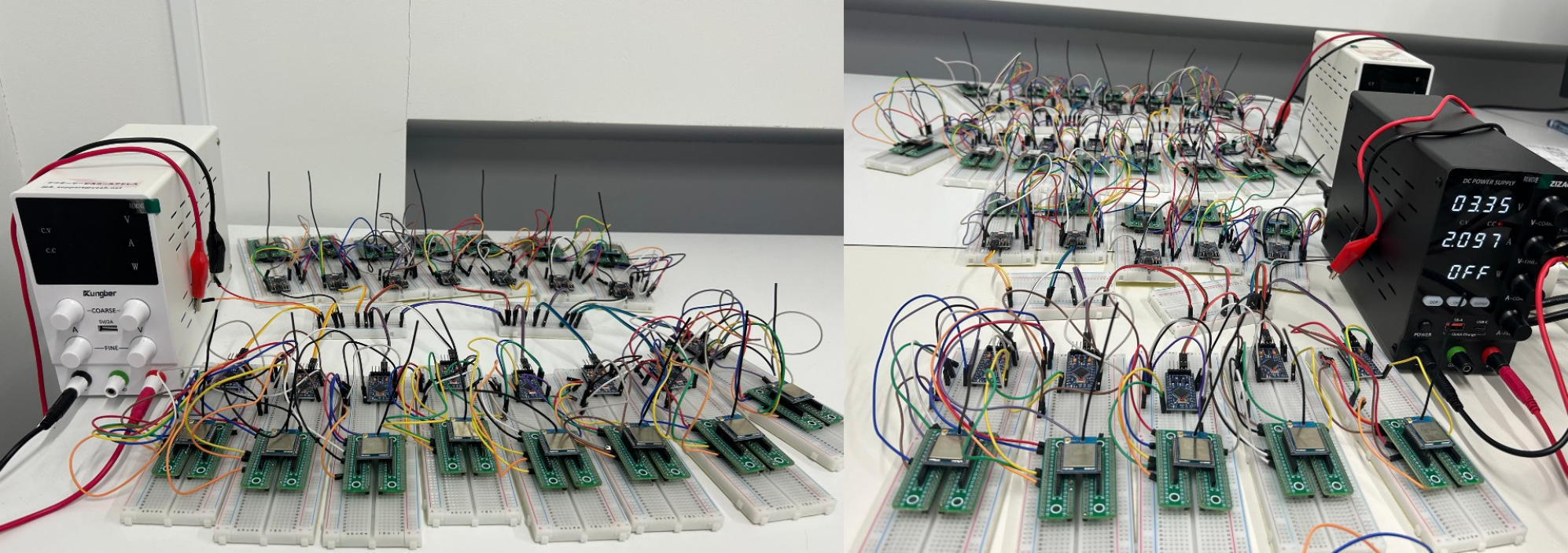}
        \caption{Transmitter.}
        \label{fig:transmitter}
    \end{subfigure}
    \hfill
    \begin{subfigure}{0.45\textwidth}
        \centering
        \includegraphics[width=\textwidth]{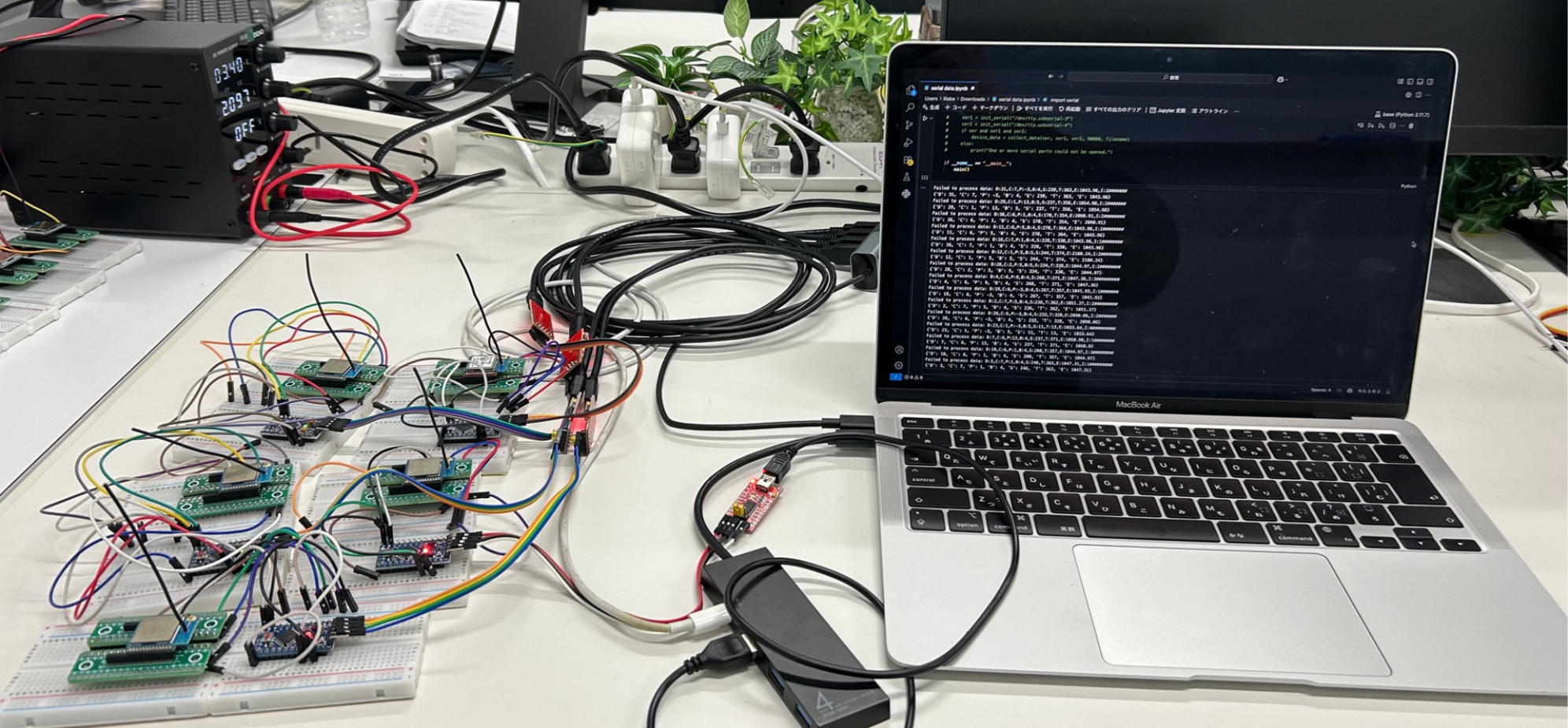}
        \caption{Receiver.}
        \label{fig:reveiver}
    \end{subfigure}

    \caption{Experimental Environment.}
    \label{fig:Experimental Environment}
\end{figure*}

Furthermore, the observed ACK result is appended to $\Omega_{c(k^*)}$ corresponding to the selected channel (line~11). 
Here, each transmission parameter set $k^*$ includes a specific channel denoted by $c(k^*) \in {C}$. Since SIC-based change detection is performed on a channel-wise basis, the ACK history is maintained per channel rather than per transmission parameter set. 
This design is motivated by the fact that environmental variations in the considered LoRa network are primarily caused by channel occupancy changes and external interference. Therefore, channel-wise ACK histories provide sufficient information for reliable change detection while keeping the memory and computational overhead low.
Using the accumulated channel-wise histories, SIC-based change detection is performed (line~12), and when the predefined detection condition is satisfied, a change in the communication environment is declared (line~13). Once a change is detected, the Beta posterior parameters of all parameter combinations are reset to their initial values, and all channel-wise history buffers are cleared (lines~14--15). This eliminates the adverse effects of outdated learning statistics and enables rapid re-adaptation to new communication environments.
The above procedure is repeated until the transmission count reaches the maximum value $T$ (lines~17--18).

\subsection{Analysis of Computational Complexity and Memory Requirements}
The computational complexity and memory requirements of the proposed and comparison methods are analyzed as follows. 
First, without SIC, both TS and UCB1-tuned require evaluating all candidate arms and selecting the best one at each transmission step. Therefore, the computational complexity is $O(|K|)$, where $|K|$ is the number of candidate transmission parameter combinations. In addition, since only the learning statistics for each arm need to be maintained, the memory requirement is also $O(|K|)$.
When SIC-based change detection is introduced, additional operations are required to maintain ACK histories and compute the corresponding statistics. Since the history length is bounded by $l_{\max}$, the additional computational complexity is $O(l_{\max})$, and the additional memory requirement is $O(N_c\times l_{\max})$.
Accordingly, the computational complexity of SIC--TS and SIC--UCB1-tuned becomes $O(|K| + l_{\max})$, and their memory requirement becomes $O(|K| + N_C\times l_{\max})$.
From the above, although the methods without SIC are the most lightweight, the additional cost introduced by SIC remains limited and preserves linear-order complexity. Therefore, the proposed method maintains its practicality for resource-constrained LoRa end devices while improving adaptability to dynamically changing communication environments.

\section{Performance Evaluation} 
To evaluate the effectiveness of the proposed method, comparative experiments are conducted against the conventional UCB1-tuned algorithm without SIC \cite{b25} and the SIC-based UCB1-tuned algorithm \cite{b32} under dynamically changing communication environments.
In all three methods, each LoRa ED autonomously and distributively selects transmission parameters consisting of a CH, TP, and BW based on the implemented reinforcement learning algorithm. The baseline UCB1-tuned algorithm selects transmission parameters using an upper confidence bound calculated from the expected reward and variance derived from ACK success statistics. The SIC--UCB1-tuned algorithm integrates SIC-based change detection into the UCB1-tuned algorithm, where environmental changes are detected from channel-wise ACK histories, and the learning statistics are reset upon detection, followed by relearning using the UCB1-tuned algorithm.

The experimental setup used in this study is illustrated in Fig.~\ref{fig:Experimental Environment}. Specifically, Fig.~2(a) provides an overview of the experimental setup, while Figs.~2(b) and~2(c) present the transmitter and receiver configurations, respectively.
Both the transmitter and the receiver were implemented using ES920LR LoRa modules and Arduino Pro Mini microcontrollers. The transmitter executes the proposed transmission parameter selection and learning process, while the receiver forwards the received data to a personal computer via serial communication for logging and analysis.

To emulate realistic non-stationary communication environments, the communication environment was dynamically varied in six phases. During transmission intervals of 1--200 and 1,001--1,200, all five channels (920.7, 921.1, 921.4, 921.6, and 921.8~MHz) were available, representing a stable communication environment. During 201--400, the 250~kHz-band channels at 920.7 and 921.1~MHz were disabled; during 401--600, the 125~kHz-band channels at 921.4 and 921.6~MHz were disabled; during 601--800, the 250~kHz-band channel at 920.7~MHz and the 125~kHz-band channel at 921.4~MHz were disabled; and during 801--1,000, the 250~kHz-band channel at 921.1~MHz and the 125~kHz-band channel at 921.6~MHz were disabled.
The environmental changes considered in this experiment emulate interference scenarios in which certain frequency bands become temporarily unavailable due to external interference sources. Such changes lead to variations in the ACK success probabilities of the affected channels, which can be identified by the SIC-based statistical change detection mechanism. 
The detection threshold was set to $\eta = 10$, which was empirically selected through preliminary experiments. A smaller threshold increases sensitivity but may cause frequent false detections, whereas a larger threshold delays adaptation to environmental changes. The selected value provides a favorable tradeoff between detection sensitivity and false-alarm robustness.
Under these dynamically varying environments, the proposed method and the comparison methods were evaluated in terms of transmission success rate and energy efficiency. Each experiment was repeated five times, and the average values were used for evaluation. The detailed experimental parameter settings are summarized in Table~\ref{tab:param_settings}. $T_{WU}$, $T_{proc}$, $T_R$ in Table ~\ref{tab:param_settings} represent the wake-up time of the LoRa device, the processing time for selecting transmission parameters by the microcontroller, and the reception time of the device, respectively.

\begin{table}[t]
\caption{Experimental Parameter Settings.}
\begin{center}
\begin{tabular}{|c|c|}
\hline
\textbf{Parameter} & \textbf{Value} \\
\hline
Number of EDs & 5, 10, 20, 30, 40 \\
\hline
CH & 920.7, 921.1 MHz (250 kHz), \\
              & 921.4, 921.6, 921.8 MHz (125 kHz) \\
\hline
TP & -3, 1, 5, 9, 13 dBm \\
\hline
SF & 7 \\
\hline
BW & 125, 250 kHz \\
\hline
Transmission Interval & 15 seconds \\
\hline
Retransmission Count & 0 \\
\hline
Number of Transmissions & 1,200 \\
\hline
Payload Length & 50 bytes \\
\hline
Startup Energy ($E_{\mathrm{WU}}$) & $56.1 \times T_{\mathrm{WU}}$ [mWh] \\
\hline
Processing Energy ($E_{\mathrm{proc}}$) & $85.8 \times T_{\mathrm{proc}}$ [mWh] \\
\hline
Reception Energy ($E_{\mathrm{R}}$) & $66 \times T_{\mathrm{R}}$ [mWh] \\
\hline
MCU Power ($P_{\mathrm{MCU}}$) & 29.7 [mW] \\
\hline
Preamble Length ($N_{\mathrm{P}}$) & 8 symbols \\
\hline
Sliding Window Length ($W$) & 10 \\
\hline
Window Shift Step ($F$) & 5 \\
\hline
ACK Sequence Length ($l_{\max}$) & 25 \\
\hline
Threshold ($\eta$) & 10 \\
\hline
$\gamma$ & 0.2 \\
\hline
\end{tabular}
\label{tab:param_settings}
\end{center}
\end{table}

\subsection{Success Rate}
\begin{figure*}[t]
    \centering
    \begin{subfigure}{0.45\textwidth}
        \centering
        \includegraphics[width=\textwidth,height=3.8cm]{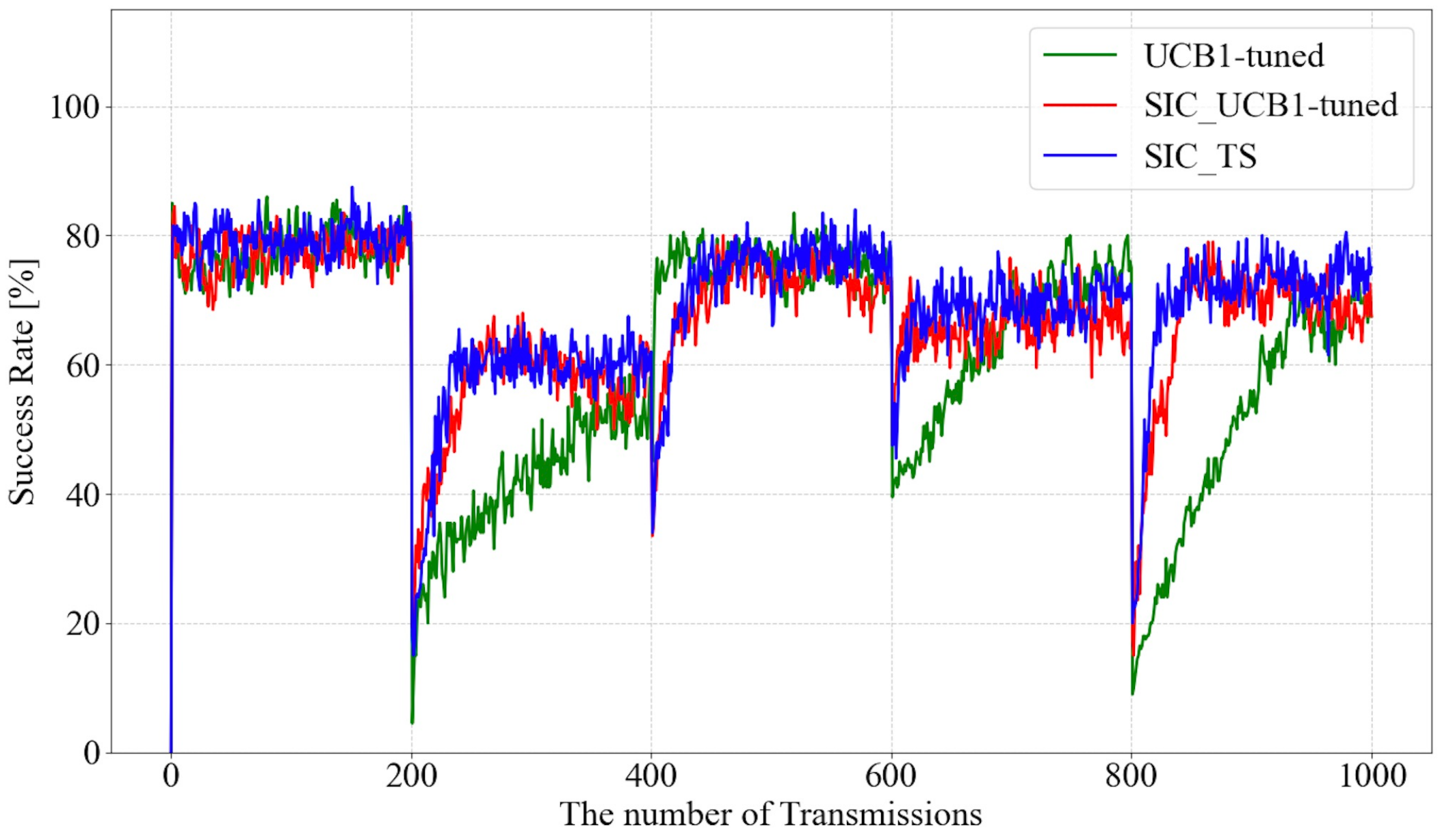}
        \caption{Transmission Success Rate vs. Number of Transmissions.}
        \label{SIC_SR}
    \end{subfigure}
    \hfill
    \begin{subfigure}{0.45\textwidth}
        \centering
        \includegraphics[width=\textwidth, height=3.8cm]{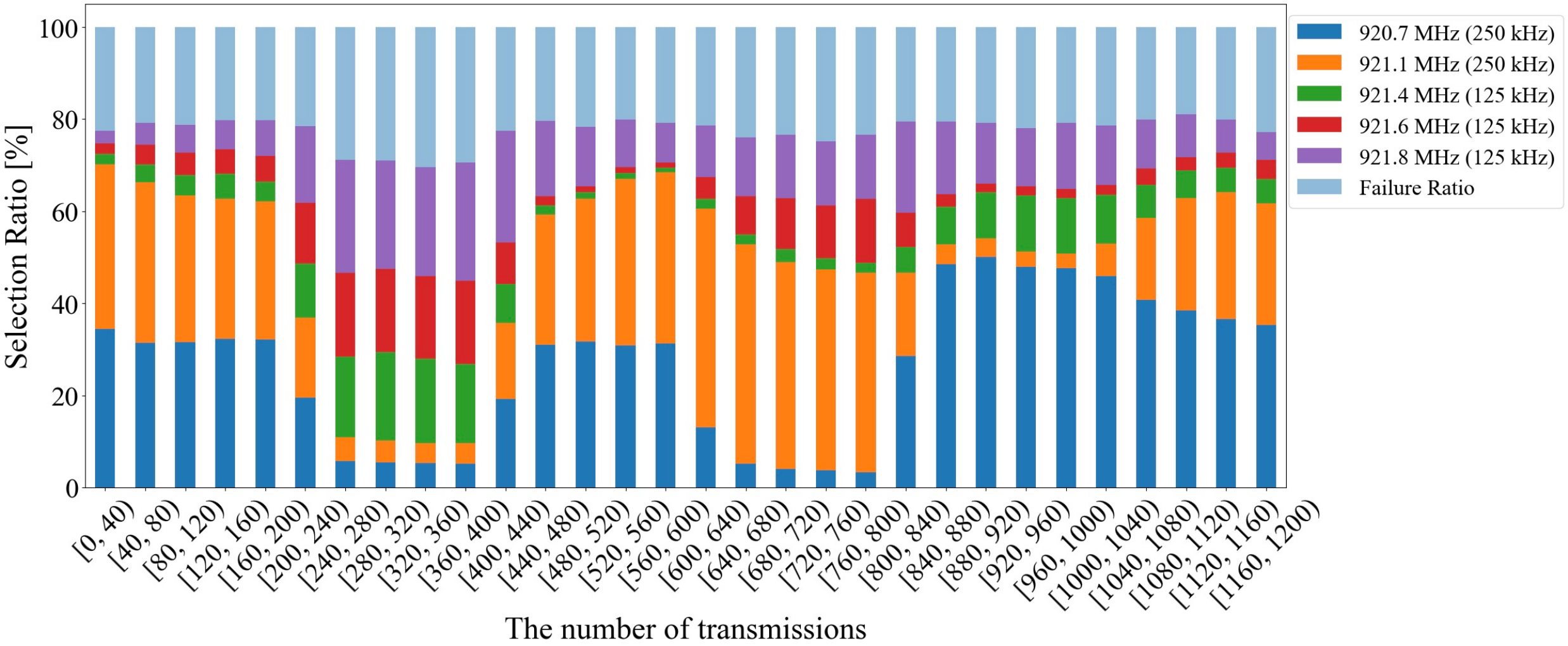}
        \caption{CH Selection Ratio (Proposed Method).}
        \label{SIC_TS_S}
    \end{subfigure}

    \vspace{0.5em} 
    \begin{subfigure}{0.45\textwidth}
        \centering
        \includegraphics[width=\textwidth,height=3.8cm]{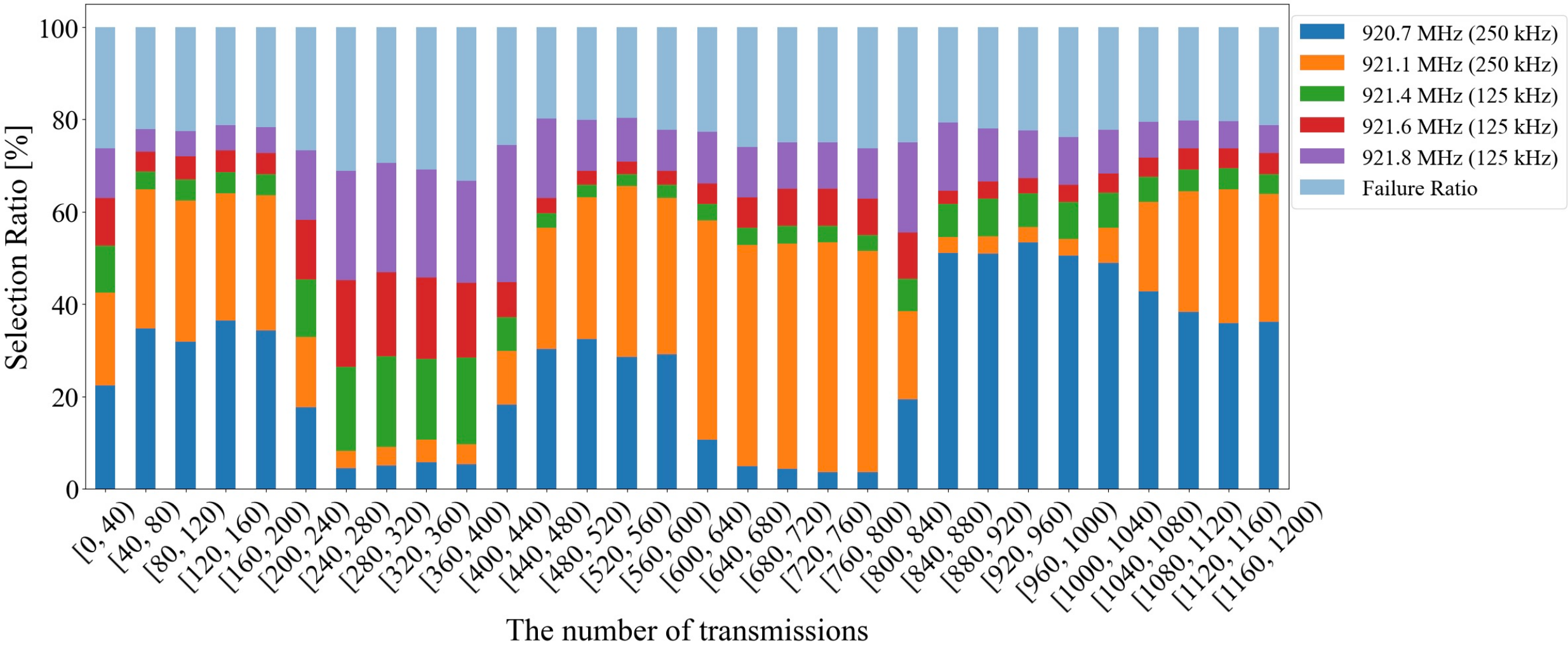}
        \caption{CH Selection Ratio (SIC-UCB1-Tuned).}
        \label{fig:sub2}
    \end{subfigure}
    \hfill
    \begin{subfigure}{0.45\textwidth}
        \centering
        \includegraphics[width=\textwidth,height=3.8cm]{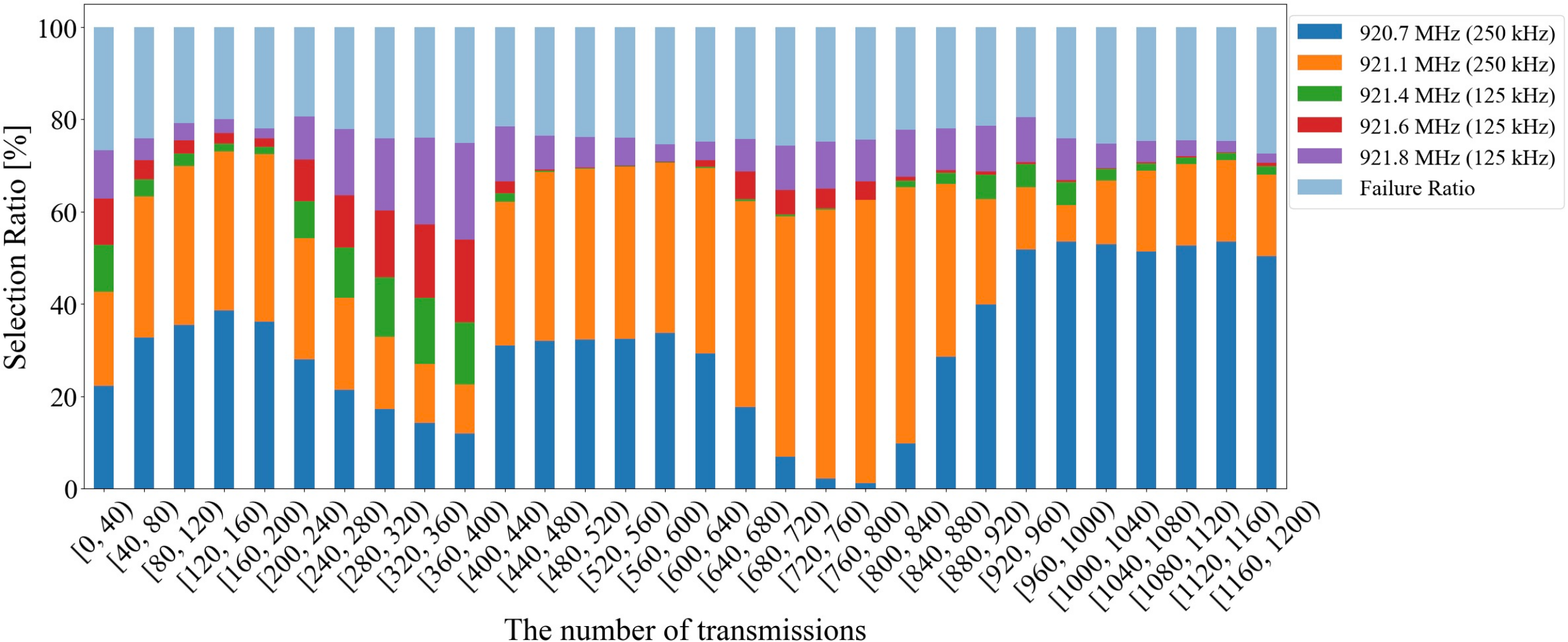}
        \caption{CH Selection Ratio (UCB1-Tuned).}
        \label{UCB_S}
    \end{subfigure}

    \caption{Success Rate and CH Selection Ratio of the Proposed and Comparison Methods.}
    \label{fig:overall}
\end{figure*}

\begin{figure*}
    \centering
    \begin{subfigure}{0.45\textwidth}
        \centering
        \includegraphics[width=\textwidth,height=3.8cm]{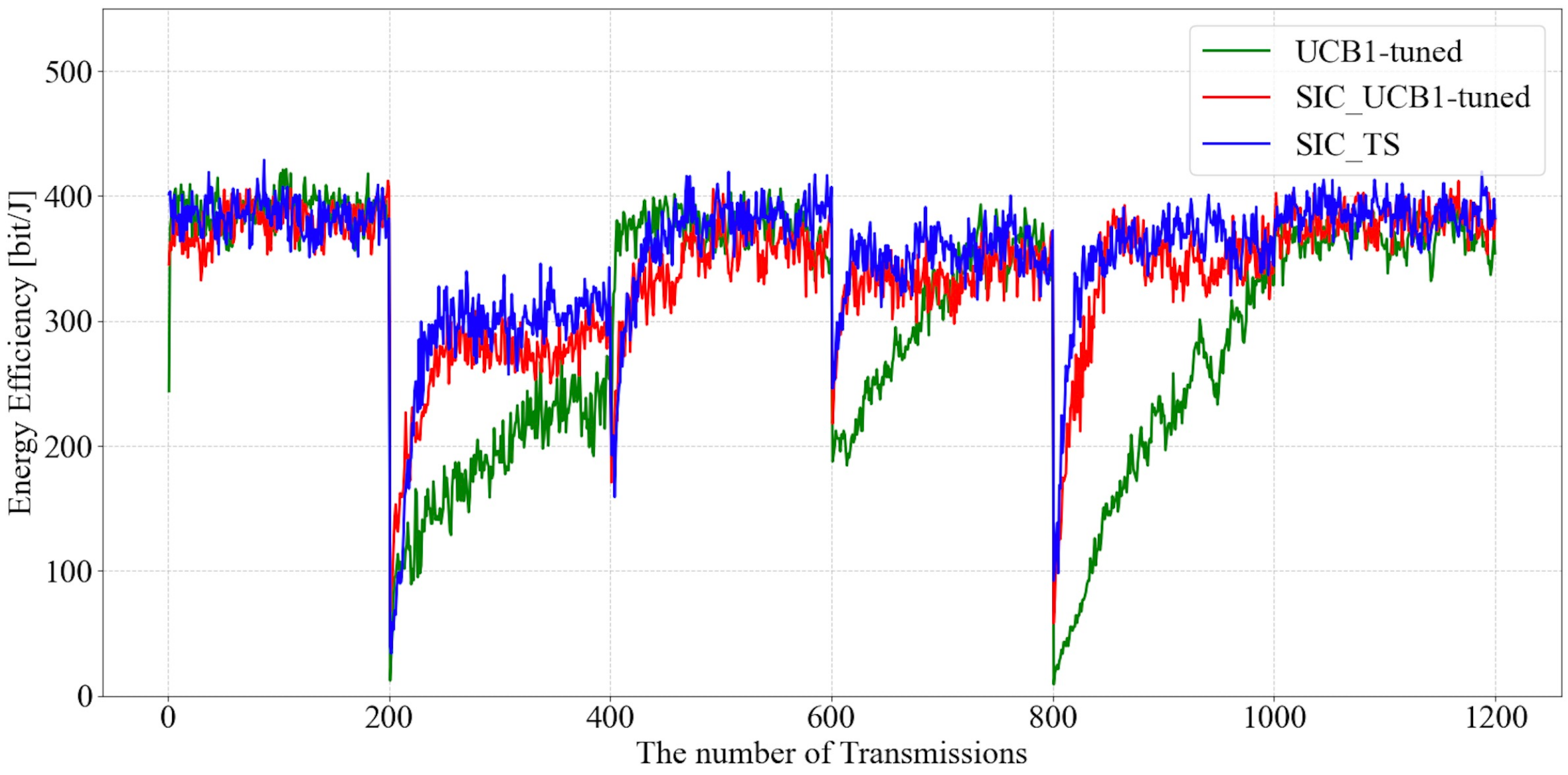}
        \caption{Energy Efficiency vs. Number of Transmissions.}
        \label{SIC_EE}
    \end{subfigure}
    \hfill
    \begin{subfigure}{0.45\textwidth}
        \centering
        \includegraphics[width=\textwidth,height=3.8cm]{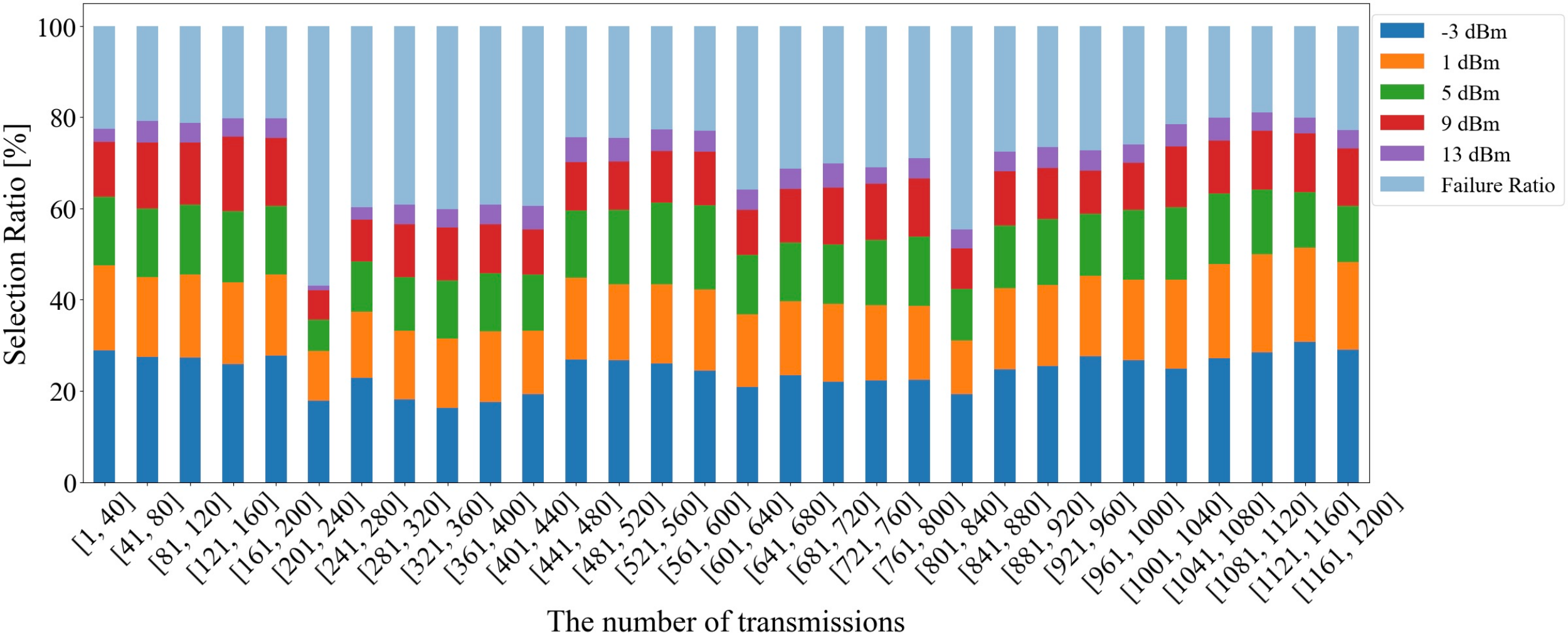}
        \caption{TP Selection Ratio (Proposed Method).}
        \label{SIC_TS_E}
    \end{subfigure}

    \vspace{0.5em} 
    \begin{subfigure}{0.45\textwidth}
        \centering
        \includegraphics[width=\textwidth,height=3.8cm]{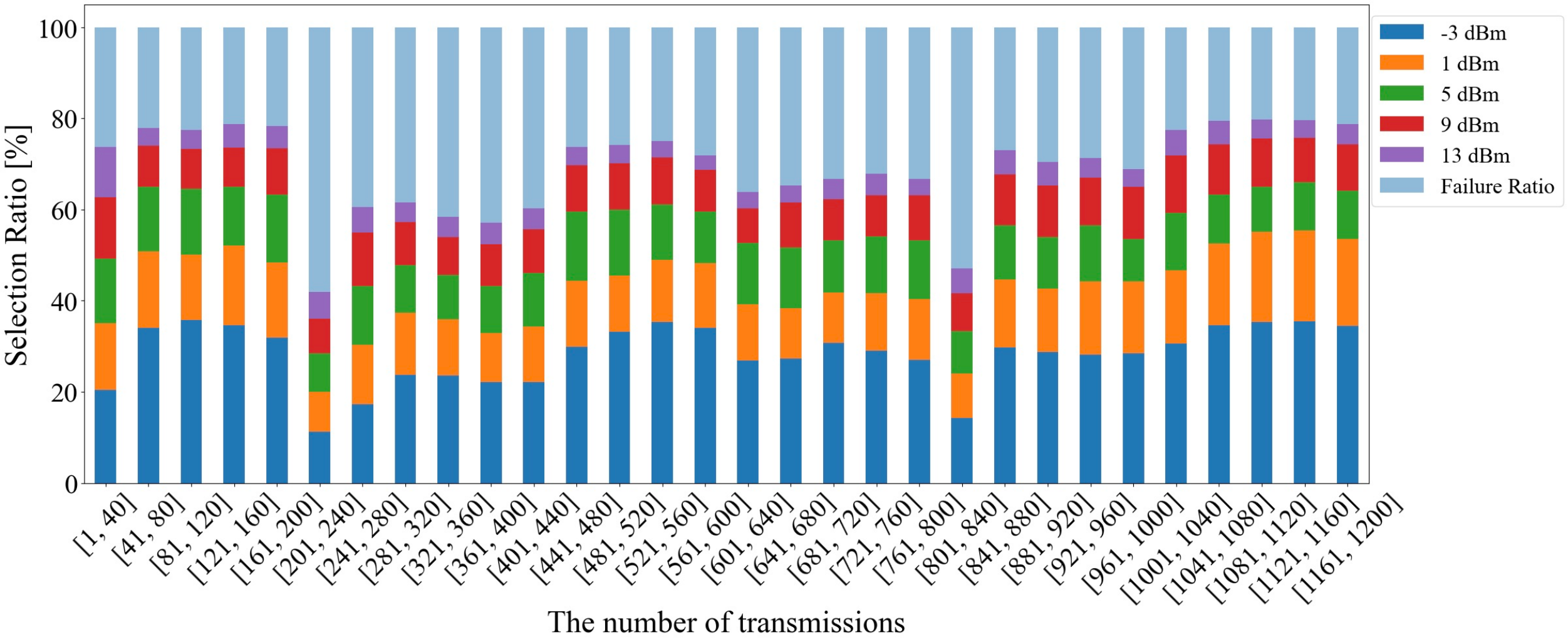}
        \caption{TP Selection Ratio (SIC-UCB1-Tuned).}
        \label{fig:sub2}
    \end{subfigure}
    \hfill
    \begin{subfigure}{0.45\textwidth}
        \centering
        \includegraphics[width=\textwidth,height=3.8cm]{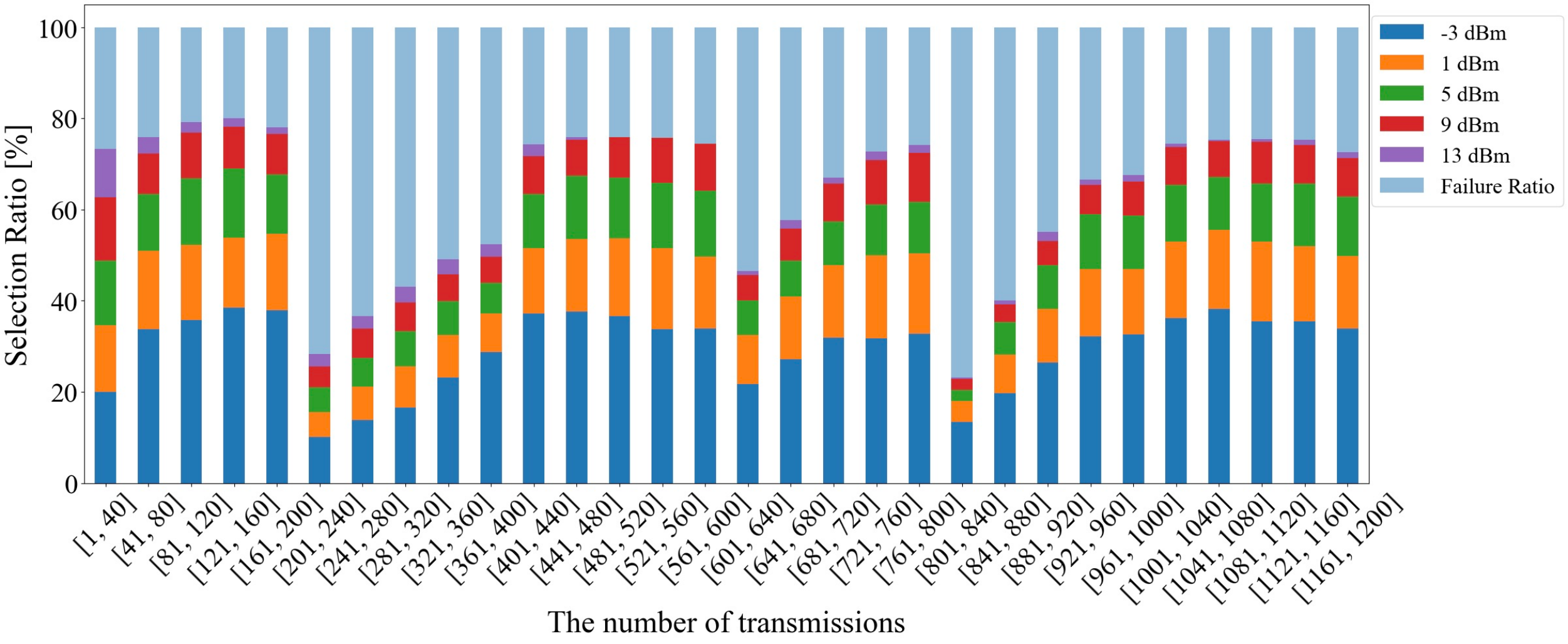}
        \caption{TP Selection Ratio (UCB1-Tuned).}
        \label{UCB_E}
    \end{subfigure}

    \caption{Energy Efficiency and TP Selection Ratio of the Proposed and Comparison Methods.}
    \label{fig:overall}
\end{figure*}
When comparing the average transmission success rates over all 1,200 transmissions under a high-density environment with 40 LoRa EDs, the proposed SIC--TS method achieves the highest success rate of 71.08\%, whereas SIC--UCB1-tuned and the conventional UCB1-tuned achieve 69.27\% and 64.00\%, respectively.
These results quantitatively demonstrate that introducing SIC significantly improves the transmission success rate and that an additional performance gain is obtained by combining SIC with TS. This performance improvement mainly originates from the ability of SIC to detect environmental changes and reset the learning statistics accordingly. In the conventional UCB1-tuned method, outdated statistics accumulated from previous environments are retained in the learning process, which delays adaptation following environmental changes. In contrast, by resetting the learning history, the SIC-based methods can rapidly adapt to new communication conditions. Furthermore, the probabilistic exploration mechanism of TS facilitates the efficient identification of promising parameter combinations, thereby providing an additional performance advantage to SIC–TS method.

As shown in Fig.~3(a), six transmission phases are configured in this experiment, where the communication environment changes stepwise at transmission counts of 201, 401, 601, and 801.
The conventional UCB1-tuned exhibits noticeable performance degradation immediately after each environmental change and requires a relatively long time to reconverge.
In contrast, both SIC--UCB1-tuned and the proposed SIC--TS maintain similarly high success rates during most transmission periods, confirming that SIC-based resetting effectively suppresses post-change performance degradation.
In particular, after the environmental changes at transmission counts of 201 and 601, the conventional UCB1-tuned suffers from prolonged performance degradation due to outdated learning statistics accumulated in previous environments, resulting in persistent inappropriate channel selections.
On the other hand, the two SIC-based methods promptly reduce the selection probability of unavailable channels and rapidly migrate to available channels, as observed in the temporal transitions of channel selection ratios shown in Figs.~3(b)--3(d).
These differences in selection behaviors directly contribute to the differences in reconvergence speed and the attainable success rates after environmental changes.
In contrast, after the environmental change at transmission count 401, the conventional UCB1-tuned temporarily converges to a high-success-rate region faster than the SIC-based methods.
This behavior is attributed to the fact that the favorable statistics learned for the 250~kHz channels during transmissions 1--200 are retained in the learning history, and the 250~kHz channels remain available during the following transmission period from 401 to 600.
Therefore, the preserved historical knowledge effectively functions in the new environment, resulting in faster reconvergence of the UCB1-tuned than that of the SIC-based methods whose learning histories are reset.

Furthermore, after the environmental change around transmission count 801, SIC--TS exhibits slightly faster reconvergence than SIC--UCB1-tuned and achieves comparable or marginally higher steady-state success rates.
This is because SIC--TS employs probabilistic TS, which preferentially explores energy-efficient parameter combinations.
However, since the number of parameter combinations considered in this experiment is relatively small, the difference in reconvergence speed between the two SIC-based methods remains limited.
When the parameter space is further expanded, for example by incorporating SF selection, the advantage of SIC--TS based on TS is expected to become more pronounced compared with UCB-based methods that require at least one exploration of all arms.

\subsection{Energy Efficiency}

Fig.~4(a) illustrates the temporal variation of energy efficiency under a high-density environment with 40 LoRa EDs. The average energy efficiency over all 1,200 transmissions is the highest for the proposed SIC--TS, achieving 328.27~bit/J, followed by SIC--UCB1-tuned with 318.65~bit/J, while the conventional UCB1-tuned achieves only 293.94~bit/J. These results quantitatively confirm that incorporating SIC significantly improves energy efficiency and that further enhancement is achieved by integrating TS.

Next, focusing on each dynamically changing period shown in Fig.~4(a), the conventional UCB1-tuned suffers a drastic degradation of energy efficiency down to below 100~bit/J around the environmental change points at approximately the 201st, 601st, and 801st transmissions, and requires a considerably long time to recover. This is because inappropriate parameter selections based on outdated learning statistics significantly reduce the transmission success rate, causing the consumed energy to fail to contribute to effective data delivery and resulting in a severe degradation of the energy efficiency. Moreover, the reconvergence to a highly energy-efficient parameter set is delayed because a large number of trials are required to identify parameter combinations that simultaneously achieve successful transmissions and low energy consumption.
In contrast, SIC--UCB1-tuned and the proposed SIC--TS exhibit much smaller degradations after environmental changes and rapidly recover to a highly efficient region exceeding 300~bit/J within a small number of transmissions.

Furthermore, by comparing the temporal variations of transmission power selection ratios shown in Figs.~4(b) to~4(d), the conventional UCB1-tuned without SIC frequently selects the minimum transmission power; however, due to its low transmission success rate, it results in the lowest energy efficiency among all methods. In contrast, while SIC--UCB1-tuned and SIC--TS exhibit similar overall tendencies, the proposed SIC--TS selects lower transmission power more frequently in the intervals [1,40], [201,240], and [801,840]. This behavior is attributed to the difference in the initial exploration mechanisms. The UCB1-tuned requires cycling through all parameter combinations after each reset, resulting in nearly uniform power selection in the early stage. On the other hand, the proposed SIC--TS incorporates an energy-consumption-based prior bias and probabilistic exploration based on TS, enabling preferential selection of low-energy parameter sets even in the initial stage. This difference in early exploration directly contributes to faster reconvergence and superior final energy efficiency after environmental changes.

\subsection{Performance Comparison for Different Numbers of LoRa EDs}

\begin{figure}
\centerline{\includegraphics[width=80mm]{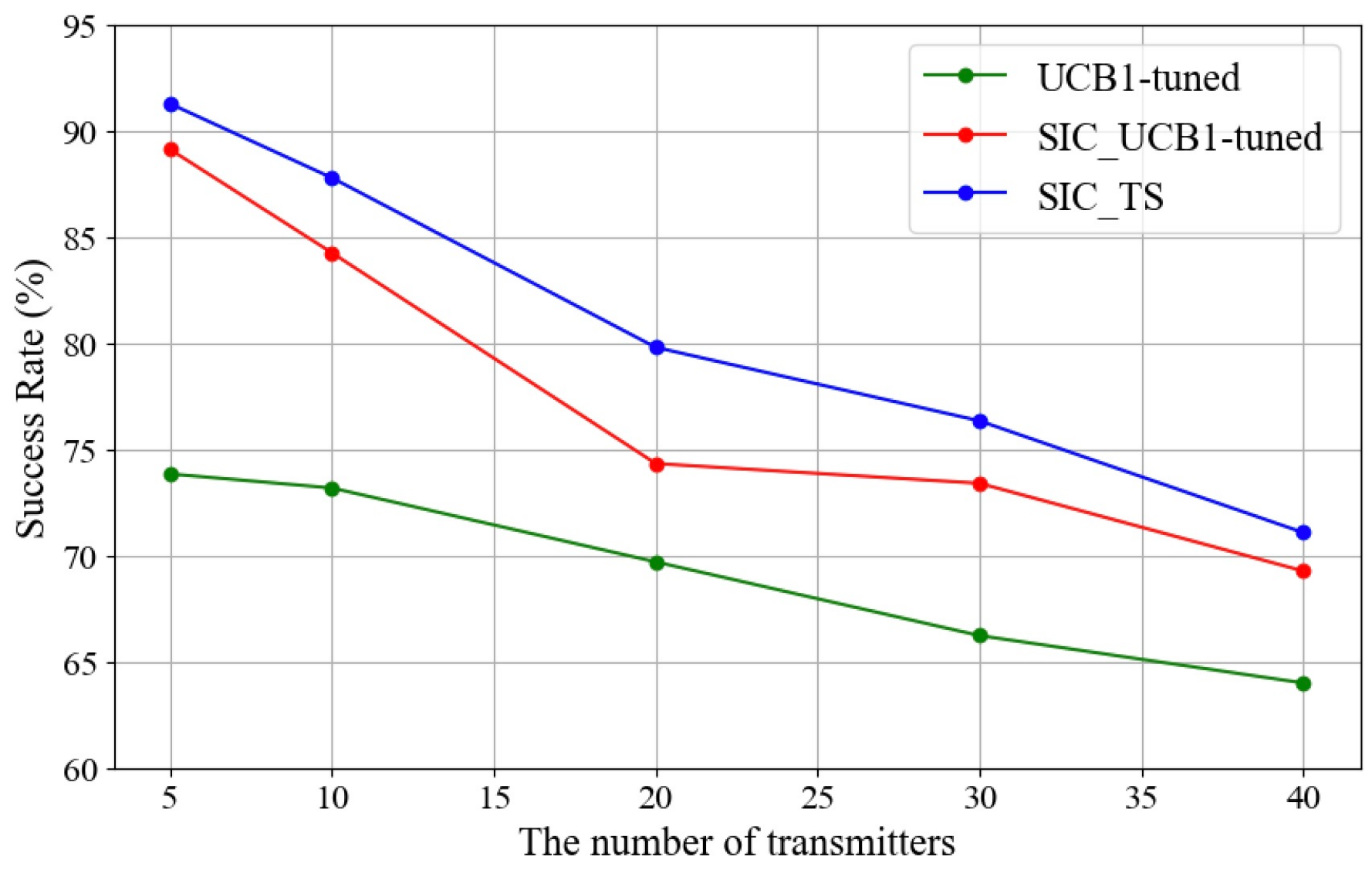}}
\caption{Success Rate vs. Number of LoRa EDs.}
\label{SR_D}
\end{figure}

Fig.~\ref{SR_D} shows the variation of the transmission success rate with respect to the number of LoRa EDs. For all methods, the success rate decreases as the number of devices increases due to the higher collision probability 
caused by network densification.
Focusing on the success rate, the proposed SIC--TS consistently outperforms both the UCB1-tuned and SIC--UCB1-tuned across all device densities. Even under high-density conditions with more than 20 devices, SIC--TS maintains approximately 10\% higher success rate than the UCB1-tuned, demonstrating robust communication reliability in congested environments. This improvement can be attributed to the probabilistic exploration of TS, which enables preferential selection of promising parameter combinations from the early learning stage, as well as the SIC-based change detection mechanism that rapidly adapts to dynamically changing wireless environments.

\begin{figure}
\centerline{\includegraphics[width=80mm]{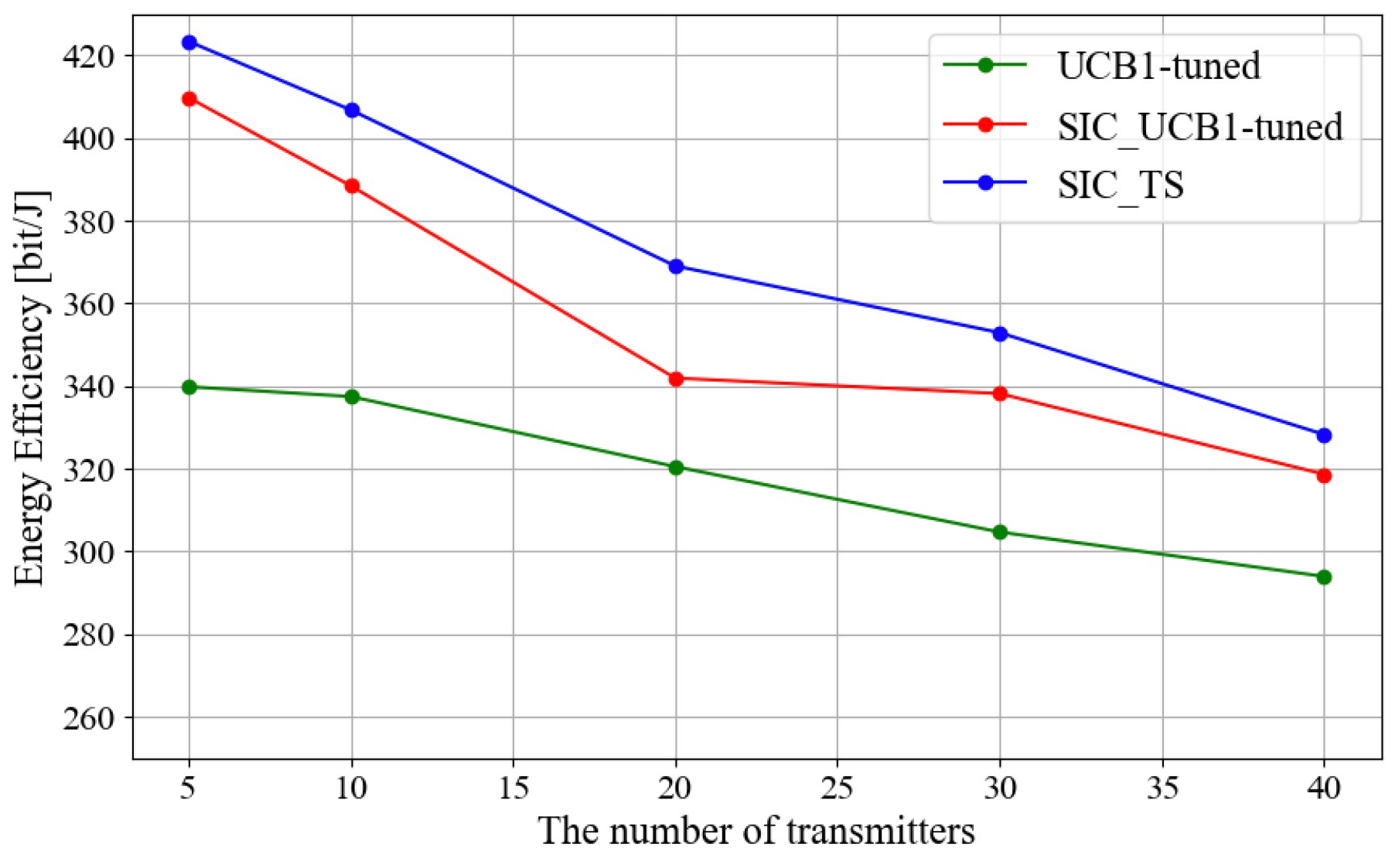}}
\caption{Energy Efficiency vs. Number of LoRa EDs.}
\label{EE_D}
\end{figure}

Fig.~\ref{EE_D} shows the variation of energy efficiency with respect to the number of transmitting devices. For all methods, energy efficiency also decreases as the number of devices increases due to the increase in channel collisions.
Regarding energy efficiency, SIC--TS achieves the highest energy efficiency across all device densities. In particular, SIC--TS maintains approximately 15--30~bit/J higher energy efficiency than the UCB1-tuned in the range of 10 to 30 devices. This indicates that the proposed method preferentially explores and selects energy-efficient transmission parameter combinations from the early learning stage. Additionally, although SIC--UCB1-tuned improves both success rate and energy efficiency compared to the UCB1-tuned by incorporating SIC-based change detection, it is consistently outperformed by SIC--TS. This is because UCB-based methods require exhaustive initial exploration over all parameter combinations, whereas TS enables efficient early-stage exploration based on probabilistic sampling from posterior distributions, resulting in faster convergence even under congested environments.

\subsection{Performance Comparison for Different Values of $\gamma$}

\begin{figure}
\centerline{\includegraphics[width=90mm]{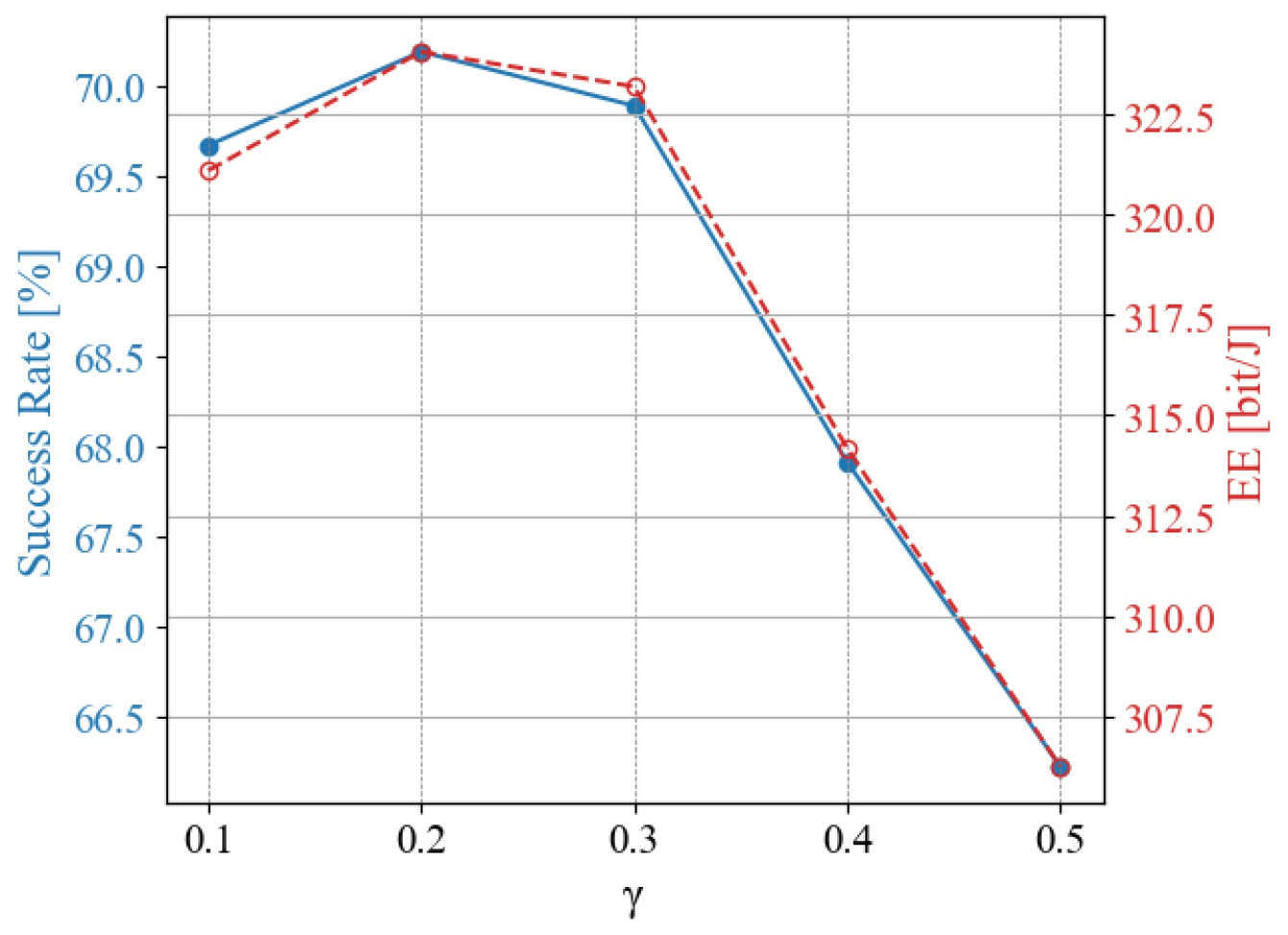}}
\caption{Success Rate and Energy Efficiency vs. $\gamma$.}
\label{ganma}
\end{figure}

Fig.~\ref{ganma} shows the variations in the transmission success rate and energy efficiency with respect to $\gamma$. In the proposed method, $\gamma$ is an important parameter that adjusts the weighting between the learning outcome based on the transmission success probability and the bias that favors low energy consumption. By changing the value of $\gamma$, the trade-off between communication reliability and energy efficiency can be controlled.
The performance evaluation results show that both the success rate and the energy efficiency achieve their maximum values when $\gamma = 0.2$, indicating the best overall performance of the system. This is because an appropriate balance is achieved between learning based on the success probability and the consideration of energy efficiency. In other words, efficient resource selection is realized by ensuring communication reliability while suppressing excessive power consumption.

On the other hand, when $\gamma = 0.4$ and $0.5$, both the success rate and the energy efficiency significantly degrade. The reason can be summarized as follows. As $\gamma$ increases, the influence of the bias term that prioritizes low energy consumption becomes dominant, making transmission configurations with insufficient transmission success probabilities more likely to be selected. As a result, the communication success rate decreases, and the amount of effectively transmitted information is reduced, leading to a deterioration in energy efficiency.
From these results, it is shown that setting $\gamma$ to an excessively large value disrupts the balance between success probability and energy efficiency, thereby causing performance degradation. Hence, setting an appropriate $\gamma$ is essential for the proposed method.

\section{Conclusion}
This paper investigates the problem of transmission parameter selection in dynamically evolving LoRa networks. The proposed fully decentralized solution integrates SIC-based change detection with TS to achieve energy-efficient transmission parameter selection.
Conventional MAB-based decentralized learning approaches rely heavily on historical observations, resulting in slow adaptation to abrupt environmental changes. The proposed method addresses this limitation by incorporating statistical change detection using SIC. This enables the timely identification of environmental shifts and adaptive resetting of learning history. This mechanism significantly improves responsiveness in dynamic network conditions.
Furthermore, the proposed framework promotes early exploration of energy-efficient transmission parameter combinations by introducing probabilistic exploration via TS together with an energy-consumption-aware prior bias. This design achieves fast convergence without requiring exhaustive initial exploration, balancing communication reliability and energy efficiency from the early learning stage.
Experimental evaluations using real LoRa devices demonstrate that the proposed method outperforms conventional UCB- and SIC-based learning approaches in communication success rate and energy efficiency under dense and dynamically changing network conditions.
Specifically, under dynamic high-density IoT deployments with 40 LoRa devices, the proposed method achieved the highest success rate of 71.08 \%, outperforming SIC-UCB1-tuned (69.27 \%) and the conventional UCB1-tuned (64.00 \%). In terms of energy efficiency, the proposed method also performed the best, achieving 328.27 bit/J, followed by SIC-UCB1-tuned with 318.65 bit/J, whereas the conventional UCB1-tuned achieved only 293.94 bit/J. Furthermore, the proposed method consistently achieved the highest success rate and energy efficiency regardless of the number of transmitters.
Overall, the results confirm that the proposed method provides a lightweight, adaptive, energy-aware, decentralized resource control solution that is well-suited for large-scale, highly dynamic IoT deployments.

% \begin{thebibliography}{00}
% \bibitem{b1} G. Eason, B. Noble, and I. N. Sneddon, ``On certain integrals of Lipschitz-Hankel type involving products of Bessel functions,'' Phil. Trans. Roy. Soc. London, vol. A247, pp. 529--551, April 1955.
% \bibitem{b2} J. Clerk Maxwell, A Treatise on Electricity and Magnetism, 3rd ed., vol. 2. Oxford: Clarendon, 1892, pp.68--73.
% \bibitem{b3} I. S. Jacobs and C. P. Bean, ``Fine particles, thin films and exchange anisotropy,'' in Magnetism, vol. III, G. T. Rado and H. Suhl, Eds. New York: Academic, 1963, pp. 271--350.
% \bibitem{b4} K. Elissa, ``Title of paper if known,'' unpublished.
% \bibitem{b5} R. Nicole, ``Title of paper with only first word capitalized,'' J. Name Stand. Abbrev., in press.
% \bibitem{b6} Y. Yorozu, M. Hirano, K. Oka, and Y. Tagawa, ``Electron spectroscopy studies on magneto-optical media and plastic substrate interface,'' IEEE Transl. J. Magn. Japan, vol. 2, pp. 740--741, August 1987 [Digests 9th Annual Conf. Magnetics Japan, p. 301, 1982].
% \bibitem{b7} M. Young, The Technical Writer's Handbook. Mill Valley, CA: University Science, 1989.
% \end{thebibliography}
\vspace{12pt}

\begin{thebibliography}{00}

\bibitem{b1}
H. Alahmadi, F. Bouabdallah, A. Al-Dubai, B. Ghaleb, A. Hussain, V. Chamola, A. Hawbani, L. Zhao, and F. R. Yu,
"A Survey on LoRaWAN MAC Schemes: From Conventional Solutions to AI-Driven Protocols,"
\textit{IEEE Commun. Surveys Tuts.}, vol. 28, pp. 2650-2690, July 2025.

\bibitem{b2}
M. Jouhari, N. Saeed, M.-S. Alouini, and E. M. Amhoud,
``A Survey on Scalable LoRaWAN for Massive IoT: Recent Advances, Potentials, and Challenges,''
\textit{IEEE Commun. Surveys Tuts.}, vol. 25, no. 3, pp. 1841--1876, May 2023.

% \bibitem{b3}
% R. Molina, J. M. Saavedra, L. F. Pedraza, and C. E. Guevara,
% ``IoT capacity building in Bolivia: learnings from a flipped and hands-on short course on LoRaWAN,''
% in \textit{Proc. 2025 IEEE Engineering Education World Conference (EDUNINE)}, Montevideo, Uruguay, 2025, pp. 1-6,

% \bibitem{b4}
% M. Cabral, A. Fuller, G. Kinyanjui and A. Lee,
% "AI-Driven Self-Optimizing Networks for Integrated LoRaWAN and 5G in Next-Generation IoT Systems,"
% in \textit{Proc. IEEE Opportunity Research Scholars Symp. (ORSS)}, Atlanta, GA, USA, 2025, pp. 1-4 doi: 10.1109/ORSS66051.2025.11121644

\bibitem{b5}
L. Aldhaheri, N. Alshehhi, I. I. J. Manzil, R. A. Khalil, S. Javaid, N. Saeed, and M.-S. Alouini,
``LoRa Communication for Agriculture 4.0: Opportunities, Challenges, and Future Directions,''
\textit{IEEE Internet Things J.}, vol. 12, no. 2, pp. 1380-1407, Jan. 2025.

\bibitem{b6}
A. Pagano, D. Croce, I. Tinnirello, and G. Vitale,
``A Survey on LoRa for Smart Agriculture: Current Trends and Future Perspectives,''
\textit{IEEE Internet of Things J.}, vol. 10, no. 4, pp. 3664-3679, Feb. 2023.

\bibitem{b7}
S. Herrería-Alonso, M. Rodríguez-Pérez, R. F. Rodríguez-Rubio and F. Pérez-Fontán,
``Improving Uplink Scalability of LoRa-Based Direct-to-Satellite IoT Networks,''
\textit{IEEE Internet Things J.}, vol.~11, no.~7, pp.~12526--12535, Apr.~2024.

\bibitem{b8}
Q. Cheng, G. Cai, J. He and G. Kaddoum,
``Design and Performance Analysis of MEC-Aided LoRa Networks With Power Control,''
\textit{IEEE Trans. Veh. Technol.}, vol.~74, no.~1, pp.~1597--1609, Jan.~2025.

\bibitem{b9}
Z. Xu, J. Luo, Z. Yin, S. Wang, C. Chen, J. Lin, R. Xiong, and T. He,
``Leveraging Imperfect-Orthogonality Aware Scheduling for High Scalability in LPWAN,''
\textit{IEEE Trans. Mobile Comput.}, vol.~23, no.~10, pp.~10111--10129, Oct.~2024.

\bibitem{b3}
N. A. Alshaer, Z. Reda, and S. A. Napoleon,
``Enhanced Adaptive Data Rate and Power Control for Resilient and Energy-Efficient LoRaWAN,''
\textit{IEEE Internet Things J.}, vol.~12, no.~22, pp.~48803--48814, Nov.~2025.

\bibitem{b4}
M. González-Palacio, D. Tobón-Vallejo, L. M. Sepúlveda-Cano, S. Rúa and L. B. Le,
``Machine-Learning-Based Combined Path Loss and Shadowing Model in LoRaWAN for Energy Efficiency Enhancement,''
\textit{IEEE Internet Things J.}, vol.~10, no.~12, pp.~10725--10739, June~2023.
\bibitem{b14}
A. Li, M. Fujisawa, I. Urabe, R. Kitagawa, S.-J. Kim and M. Hasegawa, 
"A Lightweight Decentralized Reinforcement Learning Based Channel Selection Approach for High-Density LoRaWAN," in \textit{Proc. IEEE DySPAN}, Los Angeles, CA, Dec. 2021.
% \bibitem{b7}
% Y. Sun, J. Liu, J. Yang, Q. Wu, and C. Feng,
% ``Multi-Mode Multi-Priority Low Power Wide Area Distribution Communication Network Based on LoRa,''
% in \textit{Proc. 15th Int. Conf. Communication Software and Networks (ICCSN)}, Shenyang, China, Jul.~2023, pp.~264--269.

% \bibitem{b8}
% A. Valkanis, G. A. Beletsioti, K. Kantelis, P. Nicopolitidis and G. Papadimitriou, "Balancing reliability and energy efficiency in LoRa networks using reinforcement learning,"
% in \textit{Proc. Int. Conf. Internet Technologies and Systems (CITS)}, Genoa, Italy, Jul, 2023, pp. 01-06, doi: 10.1109/CITS58301.2023.10188722.

% \bibitem{b9}
% M. A. A. Khan, H. Ma, Y. Jin, J. Ma, Z. U. Rehman and M. Rahman, 
% "Analysis of LoRa for Electronic Shelf Labels Based on Distributed Machine Learning,"
% in \textit{Proc. 42nd Chinese Control Conf. (CCC)}, Tianjin, China, Jul.~2023, pp.~3229--3234.

% \bibitem{b10}
% I. P. Manalu, S. M. Silalahi, G. I. Wowiling, M. M. T. Sigiro, E. S. Sinambela, and F. Simatupang,
% "Performance Analysis of LoRa in IoT Application of Suburban Area,"
% in \textit{Proc. 29th Int. Conf. Telecommunications (ICT)}, Toba, Indonesia, Nov.~2023, pp.~1--4.

\bibitem{d6}
A. Farhad and J.-Y. Pyun,
``AI-ERA: Artificial Intelligence-Empowered Resource Allocation for LoRa-Enabled IoT Applications,''
\textit{IEEE Trans. Ind. Informat.}, vol. 19, no. 12, pp. 11640--11652, Dec. 2023.
\bibitem{d3}
H. Yang, X. Wu, H. Ji, Z. Huang, and J. Fang,
``A Topology-Aware GNN Learning Approach for Energy Optimization in Multihop LoRa Networks,''
\textit{IEEE Internet Things J.}, vol. 12, no. 22, pp. 46596--46610, Nov. 2025.
\bibitem{d1}
Z. Lin, J. Li, H. Chen, D. Zhang, S. Gong, and B. Gu,
``Energy-Efficient Resource Allocation for Multi-Gateway LoRa Networks via Graph-Enhanced Attention Learning,''
\textit{IEEE Trans. Wireless Commun.}, vol. 25, pp. 9145-9159, Dec. 2025.
\bibitem{b11}
B. Teymuri, R. Serati, N. A. Anagnostopoulos, and M. Rasti,
``LP-MAB: Improving the Energy Efficiency of LoRaWAN Using a Reinforcement-Learning-Based Adaptive Configuration Algorithm,''
\textit{Sensors}, vol. 23, no. 4, pp. 2363, Feb. 2023.
\bibitem{d2}
Y. Guo, J. Niu, X. Zhou, T. Gu, Y. Li, and D. Fang,
``Power-Efficient Transmissions in LoRa Uplink Systems,''
\textit{IEEE Trans. Veh. Technol.}, vol. 73, no. 8, pp. 11224--11236, Aug. 2024.
\bibitem{d5}
H. Liu, L. Xiao, S. Wang, W. Lin, Z. Lv, Y. Zhan, and H. Chen,
``Learning-Based Anti-Jamming Energy-Efficient Wide-Area Communications,''
\textit{IEEE Trans. Wireless Commun.}, vol. 25, pp. 9830-9843, Dec. 2025.

%\bibitem{d4}
%M. He, M. Jin, Q. Guo, and W. Xu, ``Listen-After-Collision Mechanism for Dynamic Spectrum Access Using Deep Q-Network With an Improved Thompson Sampling Algorithm,'' \textit{IEEE Internet Things J.}, vol. 11, no. 4, pp. 6596--6606, Feb. 2024.

\bibitem{d9}
X. Zhang, Z. Lin, S. Gong, B. Gu, and D. Niyato,
``Multiagent Reinforcement Learning with an Attention Mechanism for Improving Energy Efficiency in LoRa Networks,''
in \textit{Proc. IEEE GLOBECOM}, Kuala Lumpur, Malaysia, pp. 4152--4157, Dec. 2023.

\bibitem{d10}
M. M. Salah, R. S. Saad, R. M. Zaki, K. Rabie, and B. M. ElHalawany,
``Multi-Armed Bandits for Resource Allocation in UAV-Assisted LoRa Networks,''
\textit{IEEE Internet Things Mag.}, vol. 8, no. 2, pp. 40--45, Mar. 2025.

\bibitem{b13}
I. Urabe, A. Li, M. Fujisawa, S.-J. Kim, and M. Hasegawa,
``Combinatorial MAB-Based Joint Channel and Spreading Factor Selection for LoRa Devices,''
\textit{Sensors}, vol. 23, no. 15, pp. 6687, Jul. 2023.
\bibitem{b15}
A. Li, I. Urabe, M. Fujisawa, S. Hasegawa, H. Yasuda, S.-J. Kim, and M. Hasegawa
``A Lightweight Transmission Parameter Selection Scheme Using Reinforcement Learning for LoRaWAN,''
\textit{arXiv preprint}, arXiv:2208.01824, Aug. 2022.


\bibitem{d7}
A. Scarvaglieri and F. Busacca,
``FULMINA: A Fast Multi-Armed Bandit Approach for Optimal SF Allocation in LoRa IoT Networks,''
in \textit{Proc. IEEE ICC}, Montreal, QC, Canada, pp. 5228--5233, June 2025.
\bibitem{d8}
H. Zhang, M. Li, H. Yu, H. Chen, and J. Wang,
``Dynamic Parameter Selection of LoRa Edge Nodes Using Reinforcement Learning With Link Prior Knowledge,''
\textit{IEEE Internet Things J.}, vol. 11, no. 21, pp. 34420--34433, Nov. 2024.
\bibitem{b25}
R. Ariyoshi, A. Li, M. Hasegawa, and T. Ohtsuki,
``Energy-Efficient Resource Allocation Scheme Based on Reinforcement Learning in Distributed LoRa Networks,''
\textit{Sensors}, vol. 25, no. 16, pp. 4996, Aug. 2025.

\bibitem{b16}
S. Hasegawa, R. Kitagawa, A. Li, S.-J. Kim, Y. Watanabe, Y. Shoji, and M. Hasegawa
``Multi-Armed-Bandit Based Channel Selection Algorithm for Massive Heterogeneous Internet of Things Networks,''
\textit{Appl. Sci.}, vol. 12, no. 15, pp. 7424, Jul. 2022.

\bibitem{b17}
D. Yamamoto, H. Furukawa, A. Li, Y. Ito, K. Sato, K. Oshima, S. Hasegawa, Y. Watanabe, Y. Shoji, S.-J. Kim, and M. Hasegawa,
``Performance Evaluation of Reinforcement Learning Based Distributed Channel Selection Algorithm in Massive IoT Networks,''
\textit{IEEE Access}, vol. 10, pp. 67870--67882, Jun. 2022.

\bibitem{b18}
J. Ma, S. Hasegawa, S.-J. Kim, and M. Hasegawa,
``A Reinforcement-Learning-Based Distributed Resource Selection Algorithm for Massive IoT,''
\textit{Appl. Sci.}, vol. 9, no. 18, pp. 3730, Sept. 2019.

\bibitem{b19}
M. He, M. Jin, Q. Guo and W. Xu, "Schwarz Information Criterion Based Thompson Sampling for Dynamic Spectrum Access in Non-Stationary Environment,"
\textit{IEEE Commun. Lett.}, vol. 28, no. 3, pp. 737--741, Mar. 2024.

%\bibitem{b20}
%H. Qi, F. Guo, and L. Zhu,
%``Thompson Sampling for Non-Stationary Bandit Problems,''
%\textit{Entropy}, vol. 27, no. 1, pp. 51, Jan. 2025.

% \bibitem{b20}
% Z. Shang, L. Yang, and M. Li,
% ``Two-step Thompson Sampling for Multi-Armed Bandit Problem,''
% in \textit{Proc. IEEE Chinese Control Conf. (CCC)}, Kunming, China, 2024, pp. 5802--5805.

% \bibitem{b21}
% H. Tran-Dang and D. -S. Kim,
% ``Bayesian Deep Neural Network-empowered Thompson Sampling for Context-aware Task Offloading in Dynamic Fog Computing,''
% in \textit{Proc. IEEE Int. Conf. Comput. Commun. Netw. (ICCCN)}, Tokyo, Japan, 2025, pp. 1--6.

% \bibitem{b22}
% X. Song and B. Jiang,
% ``Adaptive Satisficing Thompson-Sampling Based Bayesian Optimization for Fast Charging Design of Lithium-ion Batteries,''
% in \textit{Proc. IEEE Chinese Control Conf. (CCC)}, Kunming, China, 2024, pp. 6669--6673.

\bibitem{b23}
L. Kong, C. W. Sung, and K. W. Shum,
``Thompson Sampling and Proportional-Greedy Algorithm for Uncertain Coded Edge Computing,''
\textit{IEEE Trans. Veh. Technol.}, vol. 74, no. 3, pp. 4865--4876, Mar. 2025.

\bibitem{b24}
A. Gouverneur, B. Rodríguez-Gálvez, T. J. Oechtering, and M. Skoglund,
``An Information-Theoretic Analysis of Thompson Sampling with Infinite Action Spaces,''
in \textit{Proc. ICASSP}, Hyderabad, India, Apr. 2025.



\bibitem{b32}
R. Ariyoshi, A. Li, M. Hasegawa, M. Pan, T. Ohtsuki, and Z. Han
``Schwarz Information Criterion Aided MAB for Resource Allocation in Dynamic LoRa System,''
in \textit{Proc. INFOCOM}, Tokyo, Japan, May 2026.

\end{thebibliography}
\end{document}